\documentclass{jfm}
\usepackage{graphicx}
\usepackage{natbib}
\usepackage{psfrag}
\usepackage{bm}
\usepackage{color}
\usepackage{amsmath}
\usepackage{amsfonts}
\usepackage{amssymb}
\usepackage{natbib}
\usepackage{psfrag}

\usepackage{float}
\usepackage{pstricks}
\usepackage{color}
\usepackage{subfig}

\shorttitle{Convection in a rotating tangent cylinder}
\shortauthor{K. Aujogue, A. Poth\'erat, B. Sreenivasan and F. Debray}

\title{Experimental study of the convection in a rotating tangent cylinder}
\author{K\'elig Aujogue$^{1}$, 
  Alban Poth\'erat$^{1}$,
  Binod Sreenivasan$^{2}$
 and Fran\c cois Debray $^{3}$
}

\affiliation{
$^{1}$ Applied Mathematics Research Centre, Coventry University, Priory Street, Coventry CV15FB, UK
\\ 
$^{2}$ Centre for Earth Sciences, Indian Institute of Science, 
Bangalore 560 012, India\\
$^{3}$ Laboratoire National des Champs Magn\'etiques Intenses-Grenoble, 
CNRS/UGA-UPS-INSA, France}

\begin{document}
\maketitle

\begin{abstract}
This paper experimentally investigates the convection in a rapidly rotating 
Tangent Cylinder (TC), for Ekman numbers down to $E=3.36\times10^{-6}$. 
The apparatus consists of a hemispherical fluid vessel heated in its 
centre by a protruding heating element of cylindrical shape. 
The resulting convection that develops above the heater, i.e. within the TC, is shown 
to set in for critical Rayleigh numbers and wavenumbers respectively scaling 
as $Ra_c\sim E^{-4/3}$ and $a_c\sim E^{-1/3}$ with the Ekman number $E$. Although
 exhibiting the same exponents as for plane rotating convection, these laws 
\color{black}reflect \color{black}much larger convective plumes at onset. 
The structure and dynamics of supercritical plumes are in fact closer to those 
found in solid rotating cylinders heated from below, suggesting that the 
confinement within the TC induced by the Taylor-Proudman 
constraint influences convection in a similar way as solid walls would do. 
There is a further similarity in that the critical 
modes in the TC all exhibit a slow retrograde 
precession at onset. In supercritical regimes, the precession evolves into  
a thermal wind with a complex structure featuring retrograde rotation at high 
latitude and either prograde or retrograde rotation at low latitude
 (close to the heater), depending on the criticality and the Ekman number. 
\color{black}The intensity of the thermal wind measured by the Rossby number $Ro$ 
scales as $Ro \simeq 5.33(Ra_q^*)^{0.51}$ with the Rayleigh number based 
on the heat flux $Ra_q^*\in[10^{-9},10^{-6}]$. This scaling is in agreement with 
heuristic predictions and previous experiments where the thermal wind 
\color{black} is determined by 
the azimuthal curl of the balance between the Coriolis force and buoyancy.\color{black}
Within the range $Ra\in[2\times10^7, 10^9]$ which we explored, we also observe a 
transition in the heat transfer through the TC from a diffusivity-free regime where 
$Nu\simeq0.38E^2Ra^{1.58}$ to a rotation-independent regime where 
$Nu\simeq0.2Ra^{0.33}$.\color{black}
\end{abstract}
\section{Introduction}
\begin{figure}
\centering
\psfrag{SC}{Solid Core}
\psfrag{ICB}{Inner Core Boundary (Radius 1200 km)}
\psfrag{TC}{Tangent Cylinder}
\psfrag{LC}{Liquid Core}
\psfrag{CMB}{Core Mantle Boundary (Radius \color{green}3500 km}
\psfrag{MA}{Rocky Mantle}
\psfrag{CR}{Crust}
\psfrag{ER}{\color{red} Earth Rotation \color{black}}
\psfrag{EMF}{\color{green} Earth Magnetic Field \color{black}}
\includegraphics[width=\textwidth]{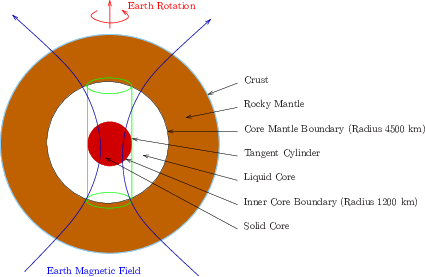}
\caption{\label{fig:tc} Schematic representation of the Earth interior. This paper focuses on the north part of the region inside the Tangent Cylinder represented in green.}
\end{figure}

This paper is concerned with convective flows confined in a cylindrical 
region by the action of the Taylor-Proudman (TP) constraint due to 
background rotation. Our prime motivation comes from the study of 
liquid planetary cores such as 
the Earth's. \color{black}{The Earth's interior is structured in layers around a solid inner core 
mostly made of iron (radius 1200 km), surrounded by a liquid core (external radius \color{black}3500 km\color{black})
where iron dominates too. A rocky mantle occupies the region surrounding the liquid core 
up to the thin crust where tectonic plates assemble (see the sketch in Fig. \ref{fig:tc}).} 
\color{black}Because of the rapid rotation of the Earth 
(Ekman numbers down to $10^{-15}$), 
the Taylor-Proudman (TP) constraint creates an imaginary boundary \color{black}{within the liquid core} 
\color{black} in the shape 
of a cylinder tangent to the solid inner core and extending up to the 
core--mantle boundary (CMB), which opposes exchange of fluid between 
regions inside and 
outside it. Consequently, the region inside the TC is subject to intense 
convection but also to an important effect of confinement of a different 
nature to that imposed by solid walls.

Until now, convection in TCs has been tackled from three different angles:
\color{black} linear stability, numerical simulations and experiments. \color{black}
Linear stability analyses mostly focused on the interplay between convection 
and rotation in plane geometries, where the TP constraint is absent. Early 
results presented in \color{black}the monograph of \cite{chandrasekhar1961hydrodynamic} \color{black}
showed that under the effect of increasing rotation, the critical Rayleigh number for 
the onset of convection increased as $Ra_c \sim E^{-4/3}$ while convective cells 
became thinner, with their wavenumber increasing as $a_{c}\sim E^{-1/3}$ (Here 
the Ekman number $E=\nu/2\Omega h^2$ is based on the background speed of 
rotation 
$\Omega$, fluid viscosity $\nu$ and the height of the fluid layer $h$). A
 notable feature of plane rotating convection is that for a Prandtl number 
above a threshold value (0.677 for free-slip boundaries), 
the unstable mode at onset is stationary \citep[see, e.g.][]{clune1993_pre}. 
Other theoretical approaches treated spherical shell geometries,
following the early study of \cite{busse1970thermal} 
but since convection outside the TC sets in at much lower 
Rayleigh numbers than inside the TC \citep{jones2007thermal}, these studies mostly 
focused on that part of the core rather than the TC.
Numerical simulations in spherical shells provided key insights into  
the global dynamics of the liquid core including the effect of the TP constraint, 
even though none was specifically dedicated to the TC. 
\cite{aubert2005steady} and 
\cite{christensen2006scaling} showed that the heat flux 
from the inner core boundary (ICB) to the CMB obeyed a somewhat universal scaling
of the form $Nu^* = 0.076(Ra^*_q)^{0.53}$ 
 when expressed in terms of the heat flux-based Rayleigh number $Ra^*_q$ and 
rotation-normalised Nusselt number $Nu^*$ (see exact definitions in equations
(\ref{eq:raq},\ref{eq:nustar})). 
A similar result was obtained for the azimuthal thermal wind, which is driven by
 latitudinal variations in the temperature perturbation
between regions inside and outside the 
TC: its intensity measured by a Rossby number built on the root
mean square (rms) velocity scaled as 
$Ro= 0.85(Ra^*_q)^{0.41}$. This scaling is consistent with 
 the hypothesis of potential vorticity conservation in a geostrophic regime
\citep{cardin1994chaotic} which 
translates into the very similar scaling of $Ro \sim (Ra^*_q)^{0.4}$. These 
scalings were obtained in dynamo simulations incorporating the full coupling 
between buoyancy, rotation and electromagnetic effects. Nevertheless, 
numerical simulations \citep{aubert2005steady} 
showed that they remained valid with or without magnetic 
field. This suggests that both the heat flux and the thermal wind are mostly 
controlled by the amount of available buoyancy. While this amount may be 
altered by electromagnetic effects, their influence becomes transparent when the 
Rayleigh and Nusselt numbers are expressed in terms of the buoyancy itself.
\color{black}
These scalings mostly concern regimes where rotation is dominant, and depend 
strongly on the level of supercriticality of convection, which controls the 
level of inertia acting on the flow, as well as the relative importance of 
rotation: where rotation does not dominate, the heat flux is expected to 
scale as in classical convection as $Nu\sim Ra^{0.25-0.43}$ in the range 
$10^5\leq Ra \leq 10^{12}$, depending on the values of $Pr$ and on whether 
the thermal or kinetic boundary layer is thickest \citep{grossmann2000_jfm}. 
According to \color{black} \cite{julien2012_prl} and \color{black} 
\cite{gastine2016_jfm}, when rotation dominates, convection near the onset 
is governed by a balance between viscous forces, buoyancy and Coriolis forces,
 whereas in strongly supercritical regimes with fast rotation, 
the diffusivity-free regime leads to a much steeper dependence of 
$Nu\sim Ra^{3/2} E^{2}$, with a transition between these regimes 
controlled by parameter $RaE^{12/7}$.
\color{black}
The transition parameters between these regimes is not specific to the TC and the question 
remains open as to whether they still stand when considering the dynamics of 
the TC only.\\
\color{black}
\cite{aurnou2003experiments} used dye visualisation to analyse convection in 
the only experiment dedicated to the TC to date. The authors identified 
convective structures, labelled "rim instabilities" that differ from those 
expected in plane rotating convection. Despite some data scattering, the 
authors suggest that the associated thermal wind was consistent with a scaling 
of the form $U \sim 2 (B/(2\Omega))^{1/2}$, where $B$ is the buoyancy flux. 
Unlike the model of \cite{cardin1994chaotic}, \color{black} this scaling reflects 
a balance between the Coriolis force and buoyancy
outside the geostrophic regime, where the excess buoyancy and the Rossby deformation radius 
are set by the balance between Coriolis and buoyancy forces \citep{maxworthy1994_jpo}. 
\color{black} Experiments 
and numerical simulations in a configuration similar to the TC show that 
this scaling is achieved in the steady state that follows the development of a
baroclinic instability at the rim of the cylinder \citep{jacobs1998_jfm,cui2001_jfm}. 
However, a characterisation of the onset of rotating convection within the 
TC is currently lacking and it is not clear how 
convection is affected by the confinement induced by the TC.\\

\cite{goldstein1993convection} and \cite{zhong1993rotating} provide an insight 
into the question of the confinement, in the configuration of a 
rotating cylinder bounded by solid 
walls. Linear stability analysis distinguishes slow and fast convective  
modes \color{black} respectively localised near the centre and the lateral wall. Their 
occurrence depends on the aspect ratio of the cylinder, the rotation and the level of 
supercriticality. A precise prediction for the onset of the wall modes was later 
derived analytically by \cite{zhang2009_jfm}. \color{black}
Importantly, in breaking translational symmetry, the presence of solid walls 
is shown to give rise to a retrograde (westward)
precession at onset that always precludes 
the occurrence of the steady modes observed in plane geometry. 
Nevertheless, the precession frequencies at onset are much lower than those 
of oscillatory convection and the associated 
precessing motion resembles more 
the retrograde thermal wind found in planetary cores than the 
travelling waves found at low Prandtl number in plane
rotating convection (\cite{chandrasekhar1961hydrodynamic,clune1993_pre}). 
Indeed, in more supercritical regimes (typically 10 times critical), 
experiments by \cite{zhong1993rotating} showed that the precession 
progressively led to a large retrograde structure
centred in the middle of the 
cylinder. \color{black} More recent studies of convection in a rotating cylinder 
focused on the turbulent regime \citep{kunnen2010_jfm}, the influence of the boundary layers
\citep{kunnen2011_jfm}, of temperature-dependent fluid properties \citep{horn2014_pf}.
\color{black}
However, it is not clear to which extent the virtual boundaries of 
the TC influence the convection in the same way as
the rigid boundaries of a 
real cylinder do; nor is it clear whether the findings of 
\cite{goldstein1993convection} and \cite{zhong1993rotating}, 
obtained for Ekman numbers of the order of 
$10^{-2}$ extend to the much lower Ekman numbers.

The purpose of this paper is to experimentally 
analyse convection in a Tangent Cylinder at low but experimentally accessible 
Ekman numbers, in view of addressing the following questions:
\begin{enumerate}
\item What are the critical conditions for the onset of convection?
\item Do the critical modes follow the phenomenology observed in
  plane or cylindrical geometries? In particular, are these modes steady?
\item What do these patterns evolve to in supercritical regimes?
\item How are heat transfer across the TC and the thermal wind affected by the 
combined influence of rotation and confinement within the TC? 
\end{enumerate}
Our experiments rely on the Little Earth Experiment (LEE) facility which we designed 
to reproduce rotating magnetoconvection in a tangent cylinder configuration 
as relevant to Earth as possible \citep{Aujogue_rsi16}. Nevertheless, all 
experiments reported in this paper were conducted in the absence of magnetic field. 
The layout of the 
paper is as follows: in section \ref{sec:apparatus}, we briefly describe LEE and
 the measurement techniques implemented in it. In section \ref{sec:patterns}, 
we analyse convective patterns in order to answer questions (a), (b) and (c). 
Section \ref{sec:heat_wind} is dedicated to the characterisation of the heat flux 
and the thermal wind while section \ref{sec:confinement} seeks to assess the 
confinement effect induced by the TC.
\section{Experimental set-up \label{sec:apparatus}}
\begin{figure}
\subfloat[Schematic of the setup]{\includegraphics[width = 0.5\textwidth]{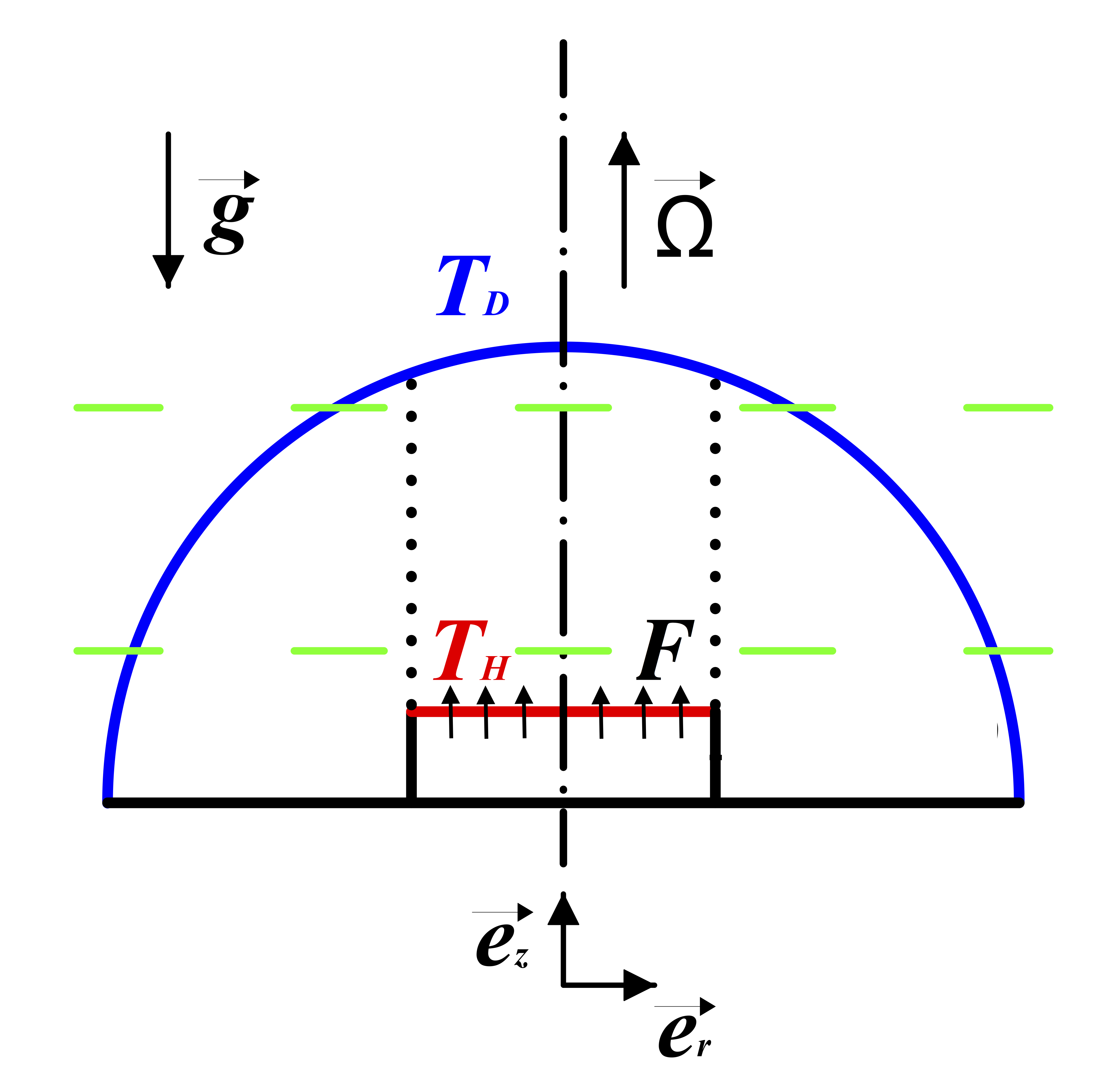}} 
\subfloat[Experimental Design]{\includegraphics[width = 0.5\textwidth]{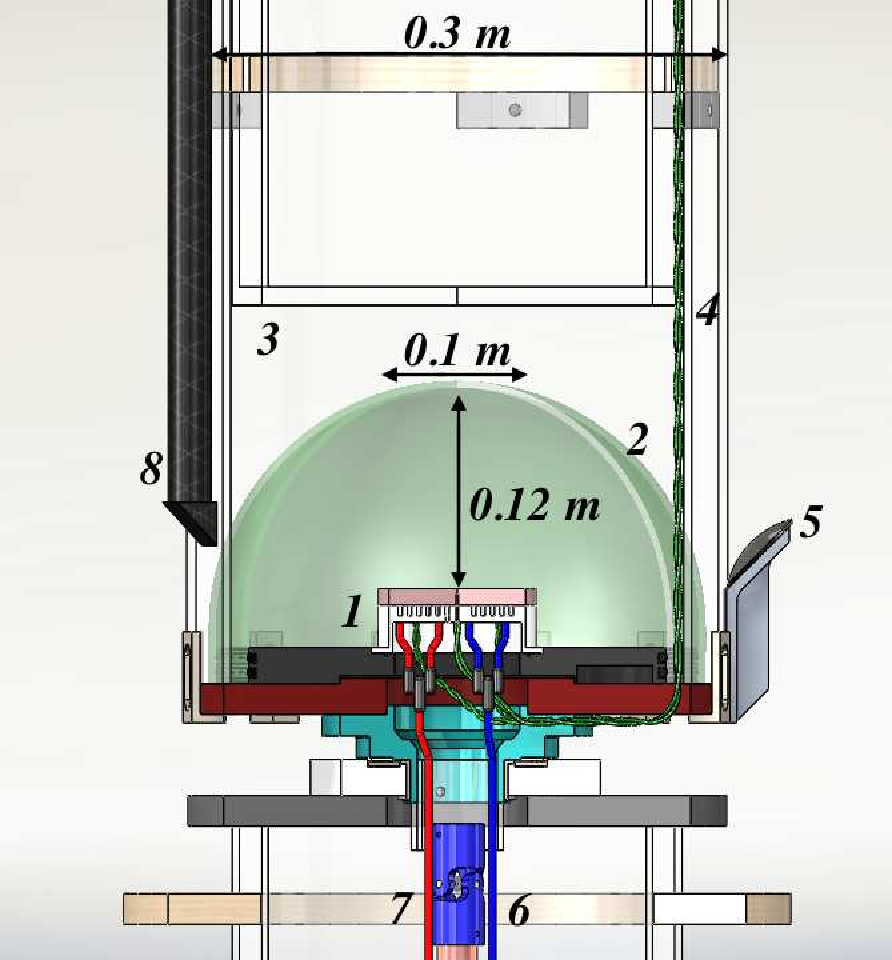}} \\
\caption{\label{fig:general} Left: vertical cross-section of the experimental 
geometry with $T_{H}$ the temperature at the heater, $T_{D}$ the temperature at the dome, $\Omega$ the 
rotation speed, $g$ being the gravity, and $F$ the heat flux through the heater. The green dashed lines 
represent the two horizontal PIV planes. The red line is the surface of the heater. The blue line is the 
surface of the dome. The dashed/dotted black line is the axis of symmetry and rotation of the experiment.
 The dotted lines are the lateral boundary of the Tangent Cylinder. Right: technical sketch of LEE, 1. 
Liquid Heater 2. Dome 3. Cooling Water 4. K-type thermocouples connected under and in the ceramic plate 
5. Mirror 6. Torque tube 7. Pipe carrying ethylene glycol 8. PIV Camera.
  (Material partially presented in \cite{Aujogue_rsi16}).}
\end{figure}

Our experimental apparatus is discussed in detail in \cite{Aujogue_rsi16}. 
Its main features are summarised in figure \ref{fig:general}.
We used a hemispherical dome of inner diameter $2R_D=276$ mm filled 
with water or sulphuric acid. At the centre of the dome, a cylindrical heater 
of height 18 mm and diameter $2\eta R_D=100$ mm protrudes
into the dome. \color{black} The cylindrical shape, with a horizontal heating surface, 
was chosen to ensure that the isothermal boundary follows an isobar when the fluid is 
at rest, as failure to observe this condition leads to spurious local baroclinic 
instabilities \citep{aurnou2003experiments}'s. \color{black} $\eta$ is the ratio of heater 
to dome radii.
Hence the thickness of the liquid corresponds to the height of
fluid above the heater at the centre of the dome $d=120$ mm. The 
heater provides an isothermal boundary condition at temperature
$T_{H}$ on its upper surface and ensures an adiabatic boundary
condition at its lateral boundary. These conditions are guaranteed
by the materials the heater is made of:
ceramic for the upper surface and polytetrafluoroethylene (PTFE)
for the lateral boundary. Indeed, considering the heat flux through
these materials ($\Phi_{\rm ceramic}$ and $\Phi_{\rm PTFE}$ respectively),
we obtained a ratio of
\begin{equation}
\frac{\Phi_{\rm PTFE}}{\Phi_{\rm ceramic}}\sim 0.0061\ll1.
\end{equation}
Therefore the flux through the lateral boundary can be neglected
with respect to the flux through the upper surface. The heater operates as a heat 
exchanger fed at a constant flow rate by a heat-carrying fluid
(Ethylene-Glycol) whose temperature is controlled in the static frame,
so the heat flux $F$ delivered to the working fluid inside the dome
is obtained by measuring the temperature difference between the inlet
and the outlet of the heater. \color{black} The ceramic the upper surface of the heater 
is made of is SHAPAL, which has a thermal conductivity of $k_S = 92$ W m$^{-1}$ K$^{-1}$. 
The corresponding Biot number, as defined by \cite{aurnou2001experiments} is 
$Bi=4.53\times 10^{-4}$, which ensures a spatial temperature inhomogeneity of less 
than 0.1\% (see \cite{Aujogue_rsi16}). The temperature is monitored during 
each experimental run and exhibits no variations greater than the measuring 
uncertainty. As a consequence, the boundary 
condition can be considered as isothermal. \color{black} Effectively, convection lowers the thermal resistance of the fluid layer by a factor $Nu$, so that when $Nu>>1$, the homogeneity of the thermal boundary condition at the heating plate is better assessed by means of a modified Biot number $Bi^{*}=Bi Nu$. The maximum value of $Nu$, we measured in all cases discussed in this paper was $Nu=155$, so that $Bi^{*}$ never exceeded $7 \times 10^{-2}$. This ensures that even with the most intense convection, the inhomogeneity in temperature distribution at the heater surface remained small. \color{black} 
The temperature at the outer surface of the dome $T_D$ is held constant
by immersing the whole system in a large volume of water. Overall, the temperature 
difference driving convection $\Delta T=T_H-T_D$ is controlled to within 
$\pm 0.2^{\circ}$C and spans values within $[0.7, 25]^{\circ}$C.
\color{black} The corresponding relative error on the Rayleigh number lies in the range 
$[1\%-28\%]$. Indeed, as the absolute error is constant the higher the temperature, the 
lower is the relative error.
\color{black}
The entire set-up \color{black} including, all PIV elements \color{black}is rotated about the 
vertical axis at angular velocity 
$\Omega$ \color{black}$\in[\pi, 3\pi]$ \color{black}rad s$^{-1}$, by means of an electric 
motor located approximately $2.5$ m below the floor of the 
hemispherical fluid domain. Even though we shall focus on rotating convection
only in this paper, it must be kept in mind that the set-up was designed 
to study both convection and magnetoconvection. Hence, this choice of 
mechanical design, which allows us to operate the setup within the bore of 
large resistive magnets whilst keeping the motor away from regions of high 
magnetic fields.
This way we aim to reproduce a geometry relevant to the 
Earth's Tangent Cylinder. 
The layout of the experiment is shown in figure \ref{fig:general}.\\
The working fluids were either water or sulfuric acid ($\rm H_{2}SO_{4}$)
of $30\%$ mass concentration and 
of respective densities $\rho_{\rm H_{2}O}=1000$ and $\rho_{\rm H_{2}SO_{4}}=1250$ kg/m$^{3}$,
viscosities $\nu_{\rm H_{2}O}=0.9\times10^{-6}$ and
$\nu_{\rm {H_{2}SO_{4}}}=2.06\times10^{-6}$ m$^{2}$/s, thermal diffusivities
$\kappa_{\rm H_{2}O}=1.4\times10^{-7}$ and $\kappa_{\rm H_{2}SO_{4}}=1.7\times10^{-7}$ m$^{2}$/s
at 20 $^\circ$C. Water was chosen for ease of use, whereas sulfuric acid was chosen
as the transparent fluid with a high electric conductivity, for experiments in 
high magnetic fields, which were conducted at the same time.
The range of parameters in which we operated the set-up is reported in
table \ref{tab:properties}. The dimensionless parameters controlling the flow
in the experiment are the Rayleigh number $Ra=g\alpha\Delta T d^{3}/\kappa$
(ratio of the buoyancy force to the viscous force),
the Ekman number $E=\nu/\Omega d^2$ (ratio of the viscous force to the Coriolis force),
the Prandtl number $Pr=\nu/\kappa$ (ratio of viscous to thermal diffusivities).\\
We measure the flow velocity with a bespoke particle image velocimetry (PIV)
system in three distinct planes: one vertical plane aligned with a
dome's diameter, and two horizontal planes positioned at $0.09$ m and
$0.02$ m above the surface of the heater (see figure \ref{fig:general}).
In the former, we measure radial and \color{black}{axial velocities} \color{black}  $u_{r}(r,z)$ and 
$u_{z}(r,z)$, where $r$ and $z$ are respectively the radial and axial 
coordinates, whereas we measure  radial and azimuthal velocities $u_r(r,\theta)$ and $u_\theta(r,\theta)$ in the latter two planes \color{black}(Here and for the remainder of the paper,
we use cylindrical coordinates with the origin at the centre
of the heater's upper circular surface). We used silver-coated particles of diameter $15\mu$m. 
\color{black} A 500 mW continuous LASER diode \color{black} generates planes of $3$ mm 
thickness and the images were acquired at $20$ frame per second with a 
camera of resolution \color{black} $1280 \times 1024$ pixels. The corresponding spatial and temporal 
resolutions are respectively $0.05$ seconds and $0.2$ mm. This system enabled us to measure velocities 
within a range of $[0.002-0.32]$m/s, with a relative error of $[5\%-10\%]$ on both velocity 
components.\color{black}\\
Temperatures are monitored in real time using 4 K-type thermocouples:
one is placed on the outer surface of the dome, one is embedded within
the top surface of the heater, and two respectively measure temperature
of the heat-carrying fluid at the heater's inlet and outlet, thus
providing accurate monitoring of $F$ (with a precision of $\pm 0.14$ W/m$^2$).
The rotating velocity is also monitored by means 
of an optical sensor. The technological details of the set-up
and validation tests are available in \cite{Aujogue_rsi16}.\\
The experimental protocol consists of 
first setting a rotation rate $\Omega$ and waiting until the flow reaches a 
solid body rotation (typically 30 min, verified by means
of PIV measurements with the heater off).
We then heat the heat-carrying fluid at a prescribed temperature,
until $\Delta T$ reaches a constant value. Only when the entire system has reached a 
statistically steady thermal and mechanical state are the output of the 
thermocouples and PIV data recorded.

\begin{table}
  \begin{center}
\def~{\hphantom{0}}
  \begin{tabular}{lccc}
  Control parameters & Water/\mbox{$\rm H_2SO_4$}  & \cite{aurnou2003experiments}  & Earth core  \\
\hline
  $E=\nu/\Omega d^2$ & $4.51\times10^{-5}$ -- $1.25\times10^{-6}$ & $9\times10^{-4}$ -- $10^{-5}$  & $10^{-15}$\\
  $Ra = g\alpha\Delta T d^3/\kappa \nu$ & $1.4\times10^{7}$ -- $2.93\times10^{9}$ & $3\times10^{6}$ -- $3\times10^{10}$ & $10^{22}$\\   
  $Pr = \nu/\kappa$ & $7/12$ & $7$  & $10^{-2}$ \\
\color{black}  $\eta = R/R_{D} $ & $0.355$ & $0.33$ & $0.35$\\
\color{black}$Fr=\Omega^2 R/g$& \color{black} 0.01--0.27 & \color{black} 0.002--0.27 & \color{black} $4\times10^{-4}$ \color{black}\\
\hline
  $\color{black}\Delta T$[$^{\circ}$C] & \color{black}[0.7-25] & \color{black}$<20$ & \color{black}$6\times 10^3$ \\
  $ \color{black}\Omega$[rad/s] & \color{black}[$\pi/2$-$4\times\pi$] & \color{black}[$0.3\times\pi$-$1.3\times\pi$] & \color{black}$7.27\times10^{-5}$ \\
  $\color{black}D$[m] & \color{black}0.1 & \color{black}0.1 & \color{black}$2.44 \times10^3$\\
\hline
  \end{tabular}
  \caption{\label{tab:properties}Range of achievable parameters in the experiment and comparison with \cite{aurnou2003experiments}'s apparatus and Earth's core parameters. Here $\nu$ is the viscosity,
    $\Omega$ the rotation rate, $d$ is the height of fluid above the centre of the heater, $g$ the gravitational constant, $\alpha$ the expansion coefficient, $\Delta T$ the temperature difference between the heater and the dome, $\kappa$ the thermal diffusivity, $D_{D}$ the diameter of the dome and $D$ the diameter of the heater. Note that the values of $Ra$ for the Earth are highly uncertain (\cite{schubert2011_pepi}).  (Material originally presented in \cite{Aujogue_rsi16}).\color{black} Values of $\Omega$, $\Delta T$ and $Fr$ were not provided in \cite{aurnou2003experiments} and have been calculated from dimensional and non-dimenisonal parameters available in this paper.\color{black}}
  \end{center}
\end{table}
\color{black}
When rotating small devices, a centrifugal acceleration can arise. If it is large 
enough to compete with gravitational acceleration, the combined acceleration is not 
vertical anymore and the resulting convection patterns can be significantly altered. 
The Froude number $Fr=\Omega^2 R/g$ represents the ratio of the centrifugal acceleration to gravitational acceleration. 
For $Fr<0.4$, convection in a solid cylinder of aspect ratio 1 at and $Pr=7$, (close 
to our configuration) is not significantly different to convection at $Fr=0$: 
it sets in via a pitchfork bifurcation leading to the appearance of wall modes.
$Fr$ then acts as a small imperfection parameter softening the bifurcation and slightly 
increases the critical Rayleigh number of every wall mode \citep{marques2007_jfm}. 
These observations were confirmed in a cylinder of larger aspect ratio of 4 
\citep{curbelo2014_pre}. For larger values of $Fr$, there is no base conduction state 
anymore and a large axisymmetric circulation driven by the centrifugal force exists, 
that is stable to three-dimensional perturbations. Whilst $Fr$ is very small for 
planets, it is in the range $0.01-0.27$ in LEE. In this range, the convection is not 
expected to be structurally altered, and the centrifugal force may have a small 
stabilising influence on the onset of wall modes. In all experiments, we observed 
a steady base state without any evidence of the axisymmetric recirculation that 
would occur at higher Froude numbers, which confirms that the influence of 
centrifugal acceleration remained weak. \color{black}
\section{Structure of the convective patterns \label{sec:patterns}}
\subsection{Onset of convection}
\begin{figure}
  \centerline{\includegraphics[width=1\textwidth]{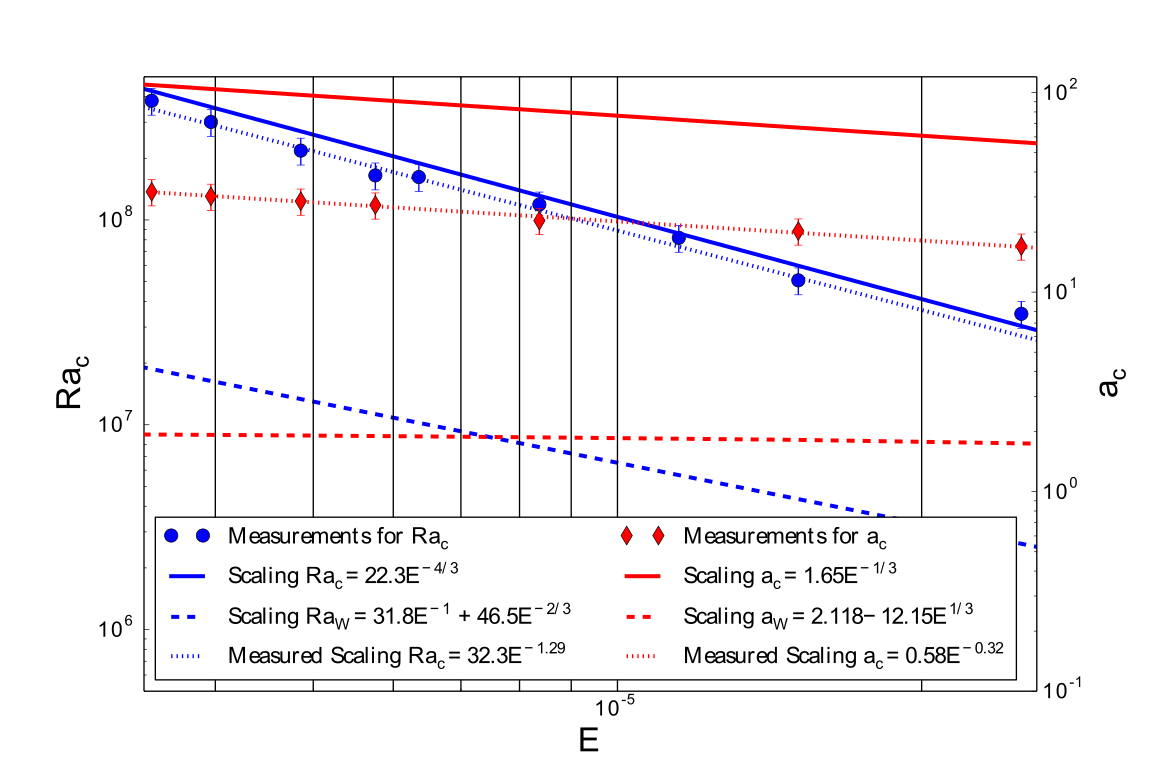}}
  \caption{\label{fig:onset} Critical Rayleigh number $Ra_{c}$ and
    critical wave number $a_{c}=2\pi R/r_0$ at onset \emph{vs.} the Ekman number $E$ \color{black} for three geometries: present case (symbols and dotted lines), infinite plane layer (solid lines) and 
wall modes in a finite cylinder obtained from \cite{zhang2009_jfm}'s theory ( Eqs. (\ref{eq:raw} and \ref{eq:aw}), dashed lines).\color{black} }
\end{figure}

\begin{figure}
  \subfloat[$E=2.51\times 10^{-5}$ and $Ra=3.48\times10^{7}$]
           {\includegraphics[width = 0.5\textwidth]{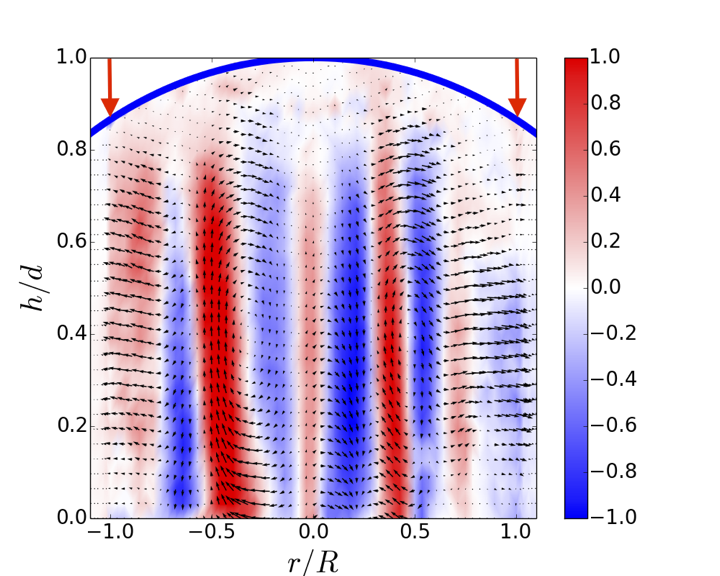}} 
           \subfloat[$E=3.36\times 10^{-6}$ and $Ra=3.82\times10^{8}$]
                    {\includegraphics[width = 0.5\textwidth]{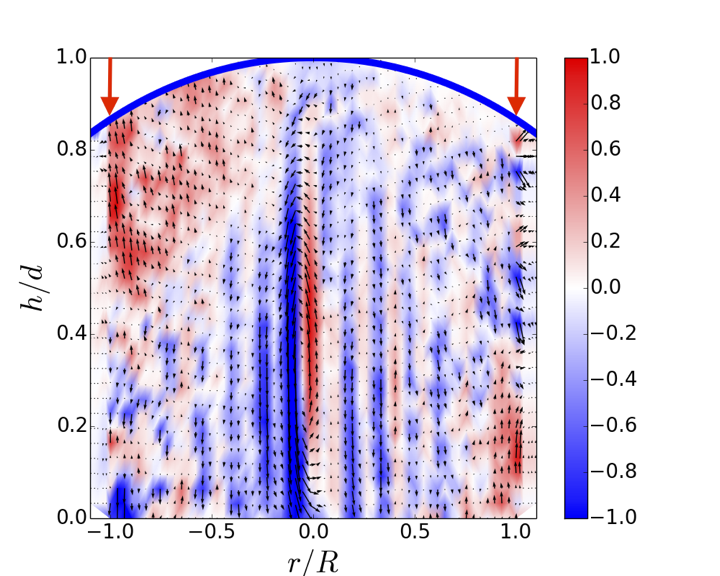}} \\
\caption{\label{fig:vert} Average over time of the vertical component of 
velocity from PIV measurement above the liquid heater \color{black}at the onset of 
convection. \color{black}The averaging time is 15 times smaller than the precession 
timescale $\tau_{p}=2\pi/\omega_p$.  
Velocities are normalised by the maximum velocity of the flow.
Here $z$ and $r$ are normalised by the maximum height above the heater
and the radius of the heater respectively.
The blue line represents the inner boundary of the glass dome. \color{black}The red arrows point to the position of the TC.}
\end{figure}

In this section, we focus on the flow near the onset of convection. 
The combined effect of buoyancy and rotation generates a flow structured in 
columns at the onset. This is illustrated by the contours of vertical velocity 
obtained for different values of $E$ on Fig. \ref{fig:vert}. Note that the columns 
are exclusively confined within the TC. \color{black}This can be inferred 
from the sharp drop of vertical velocity visible all along the TC boundary. It can also be 
seen from the radial profiles of velocity in Figs. \ref{fig:cf_a} and \ref{fig:cf_c}. 
It is noteworthy that the configuration of our experiment favours confinement within 
the TC because (i) the heat flux is generated along the whole horizontal section of the TC and nowhere 
outside it and (ii) gravity is always parallel to the TC boundary, unlike in the Earth 
where they intersect. Hence, the buoyancy force never directly acts as to break the TP 
constraint in LEE, unlike in the Earth. However, the small residual flow 
outside the TC is further attenuated by an increase in background rotation (see fig 
\ref{fig:cf_a}-\ref{fig:cf_d}). Conversely, qualitative experiments conducted without 
rotation show very different picture where plumes take the entire space above the heater,
for the flow to return entirely outside of the would-be TC.
Considering the sharpness of the velocity 
jump across the TC boundary, these elements give a good evidence that the confinement we 
observe mainly results from TP contraint. \color{black}
Columns become thinner at lower values of $E$, a tendency that has been 
previously observed both in spherical and plane
geometries  
\citep{sreenivasan2006azimuthal,chandrasekhar1961hydrodynamic,Aujogue_pof14}.\\
On figure \ref{fig:onset}, we show that variations of the critical Rayleigh 
number with $E$ follow a scaling of $Ra_c=(32.3\pm 4)\times E^{- 1.29 \pm 0.05}$.
Despite a very different geometry, this scaling is in 
good agreement with the theoretical prediction of  $Ra_c= 22.3\times E^{-4/3}$ for 
rotating convection in an infinitely extended plane layer
\citep{chandrasekhar1961hydrodynamic}.
This result suggests that the critical Rayleigh number is not
measurably affected by the geometry \color{black}differences between a plane and our TC
and reflects a very robust feature of plane rotating convection.\\ \color{black}
We shall now analyse the horizontal size of the convective structures present 
at the onset. Since translational invariance in the horizontal plane is lost in 
our geometry, we extract the horizontal size of convective structures
by seeking the separation $r_0$ corresponding to the first zero of the spatial 
correlation function built from $u_{r}$, and averaged over time and $z$:
\begin{equation}
C_{u_{r}}(\delta r)=\left\langle \int_\mathcal V u_{r}(r+\delta r)u_{r}(r)drdz\right\rangle_t, 
\end{equation}
where $\mathcal V$ represents the intersection of the meridional plane
lit by the PIV laser and the \color{black} region inside the \color{black} TC. 
At the onset, the associated wavenumber $a_c=2\pi R/r_0$ can be compared to the 
critical wavenumber predicted for the onset of rotating convection in an 
infinite plane layer (Fig. \ref{fig:onset}). 
In the TC geometry, we find a scaling of 
\begin{equation}
a_{c}=(0.58\pm0.08)\times E^{-0.32\pm0.05}
\end{equation}
when the plane layer theory predicts $a_{c}=1.65\times E^{-1/3}$ 
\citep{Aujogue_pof14}. 
Although both structure sizes exhibit the same scaling exponent, critical wave 
numbers are significantly lower in the TC geometry than for the 
infinite plane layer. We shall see in section \ref{sec:supercritical} that the 
reason for this discrepancy originates in the topological structure of the 
critical modes.

Lastly, a remarkable feature of the onset of rotating convection in plane 
layers is that for the values of $Pr$ considered in this paper, linear \color{black}stability 
\color{black} 
predicts a steady  critical mode \citep{clune1993_pre}. By contrast, in all our 
measurements, we found a time-dependent flow at the onset. Inspection of the flow in horizontal 
planes reveals that the convective plumes are subject to a slow retrograde 
precession. The variations of the corresponding frequency $\omega_p$ derived 
from the maximum velocity along $\theta$ in the horizontal plane
and normalised by the background angular frequency
are reported in figure \ref{fig:wave}.
The precession at the onset of convection and beyond has been studied 
theoretically and experimentally in \cite{goldstein1993convection} and 
\cite{zhong1993rotating} in rigid rotating cylinders of various aspect ratios. 
In these studies, the authors showed that the loss of translational 
symmetry in the radial direction necessarily induced a precession in the 
critical mode and that the corresponding frequency normalised by the 
background rotation $\Omega$, $\omega_{p_0}$ obeyed a 
scaling of the form $\omega_{p_0}=\delta E^{-1}$. 
Experiments by \cite{ecke1992hopf} determined a value of  $\delta=0.1$ 
for a radius-to-height aspect ratio of $\Gamma=1$. 
Our measurements produce a value of $\delta = 0.07 \pm 0.005$.
This value is close to that found by \cite{ecke1992hopf},
despite a lower aspect ratio 
(measured at the centre of the heater) of $\Gamma=R/d=5/12$.\\  
\color{black} The precession of convective structures at the onset in a cylindrical geometries is 
normally associated with the onset of modes that are localised either at the centre of 
the cylinder or near the wall, respectively centre modes and wall modes \citep{goldstein1993convection}. 
The values of the precessing frequencies and of $\delta$ we find point to wall modes rather than centre 
modes \citep{ecke1992hopf}. 
Nevertheless, in cylinders bounded by solid walls, the onset of wall modes normally takes places 
at a lower Rayleigh number (denoted $Ra_W$) than the steady modes that ignites plane convection 
for the same value of $E$. \cite{zhang2009_jfm} provide an estimate for $Ra_W$ and the corresponding 
azimuthal wavenumber $a_W$, which are reported on Fig. \ref{fig:onset} and expressed in our 
notations and for the present configuration as
\begin{eqnarray}
Ra_W&=& 31.81 E^{-1}+46.49 E^{-2/3},\label{eq:raw}\\ 
a_W&=& 2.118-12.15E^{1/3} \label{eq:aw}.
\end{eqnarray}
The fact that in the TC, $Ra_c$ follows the scaling for plane convection rather than that 
of wall modes is perhaps explained by the fact that unlike a solid wall, 
the Taylor-Proudman constraint is absent when the fluid is at rest. 
\color{black}
Wall modes would be expected to occur at a lower critical Rayleigh number than the onset of 
plane convection. Since, however, no confinement exists at such Rayleigh numbers, the mechanism for 
their onset is absent. On the other hand, as soon as convection starts, \emph{i.e.} for Rayleigh 
numbers close to those for which the plane layer would be unstable, the TP constraint becomes 
active and selects modes that are closer to those observed in a cylinder than those observed between infinite planes.
Hence, the TP constraint may  
not influence the flow right at the onset, but only once unstable structure have sufficiently 
developped.\color{black}\\
In summary, the study of critical Rayleigh, plume size, and precession 
frequency  shows that the onset of convection in a TC corresponds 
to a hybrid behaviour between those of convection in an infinite plane layer 
and in a rotating cylinder. Certainly, the scaling for the critical Rayleigh 
number from the plane layer theory is reproduced in the TC but 
size and time-dependence of the flow are to some extent more accurately described by the 
phenomenology of convection in rigid rotating cylinders. 

\subsection{Supercritical flow patterns \label{sec:supercritical}}
\begin{figure}
	\subfloat[$R_{c} = 0.13$, \color{black}fast wall mode (WM) at onset.]{\includegraphics[width = 0.5\textwidth]{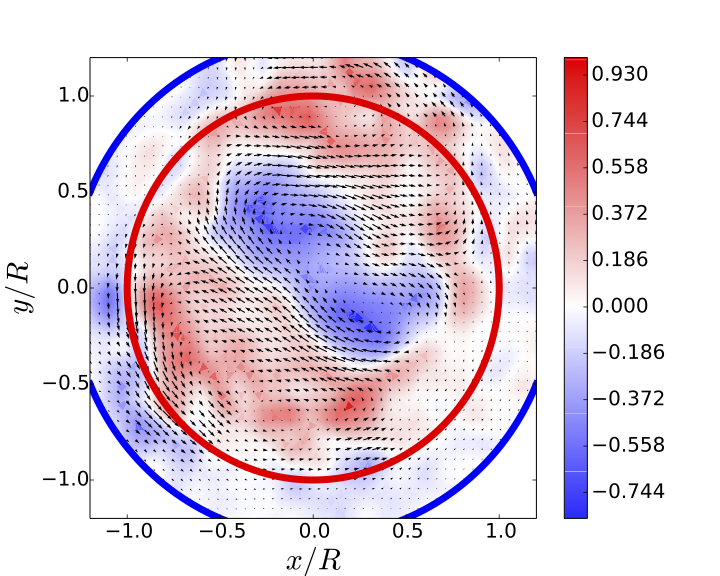}\label{fig:4a}} 
	\subfloat[$R_{c} =  1.26$, \color{black}quasi-geostrophic (QG)  mode.]{\includegraphics[width = 0.5\textwidth]{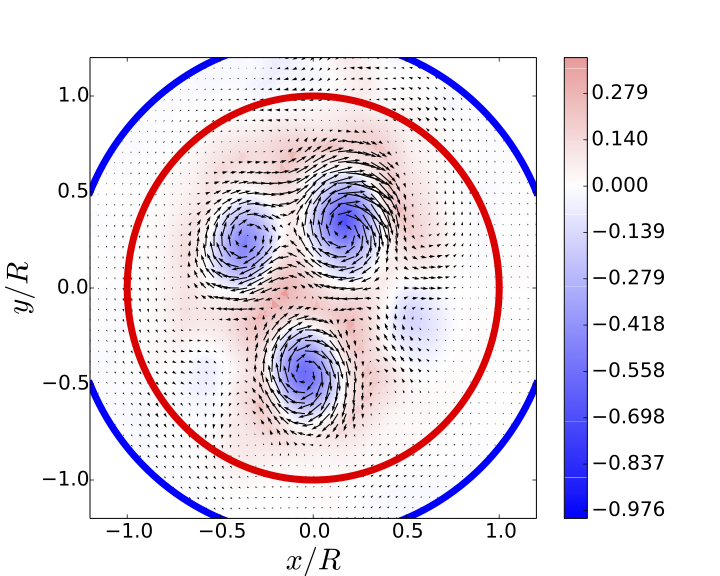}\label{fig:4b}} \\
	\subfloat[$R_{c} =  3.58$, \color{black}QG and WM interaction.]{\includegraphics[width = 0.5\textwidth]{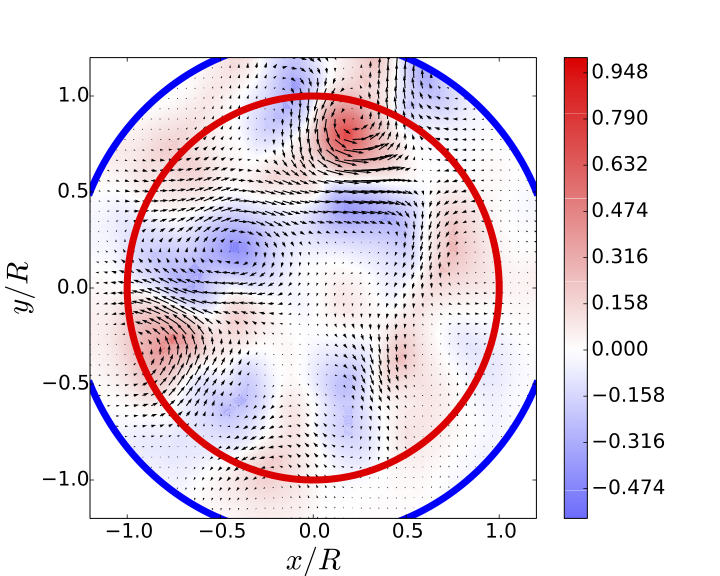}\label{fig:4c}} 
	\subfloat[$R_{c} =  6.29$, \color{black}merging of centre modes.]{\includegraphics[width = 0.5\textwidth]{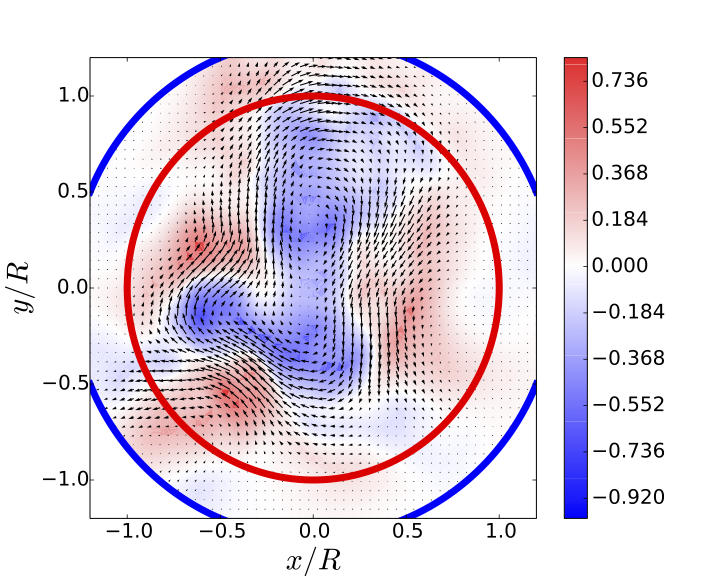}\label{fig:4d}} \\
	\subfloat[$R_{c} =  8.25$, \color{black}merging of centre modes.]{\includegraphics[width = 0.5\textwidth]{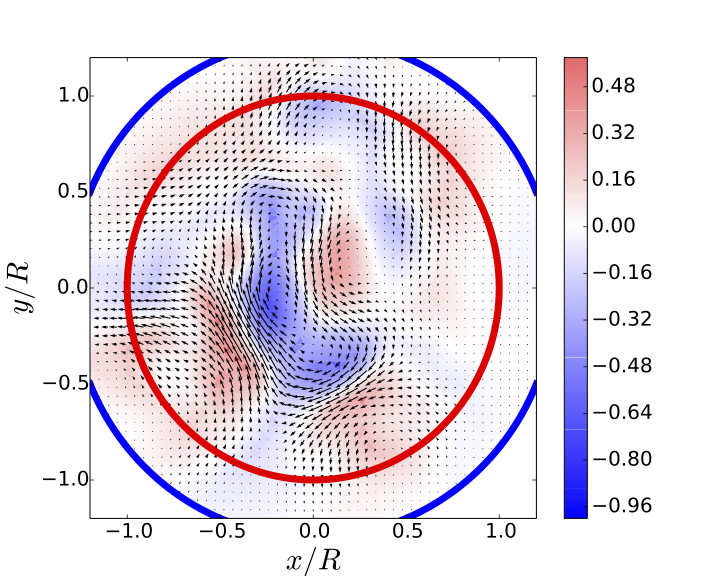}\label{fig:4e}} 
	\subfloat[$R_{c} =  11.27$, \color{black}central retrograde vortex.]{\includegraphics[width = 0.5\textwidth]{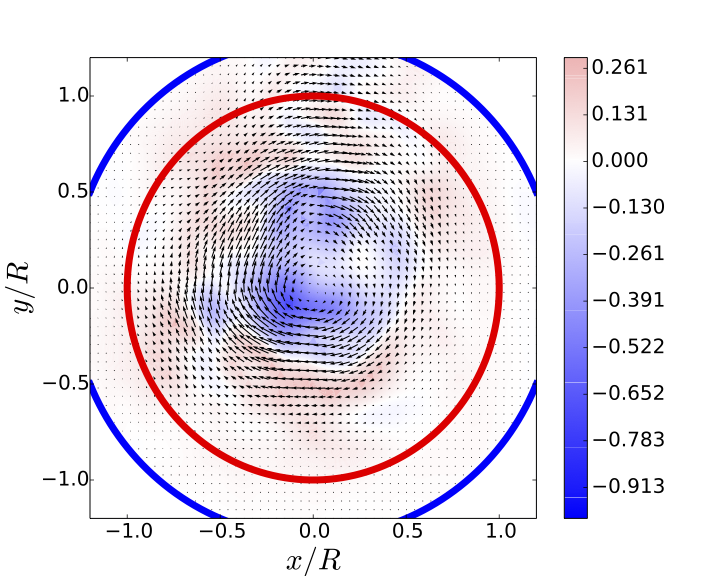}\label{fig:4f}} \\
	\caption{\label{fig:hor1}Time averaged velocity fields (arrows)
          with vorticity field (colorbar) in the horizontal plane at $z/d=3/4$ for
          different Rayleigh numbers with $E=1.15\times 10^{-5}$.
          The blue line represents the boundary of the glass dome.
          The red line represents the position of the heater that
          defines the TC. \color{black} Wall modes can be identified where vorticity 
extrema are present near the boundary of the TC (\emph{e.g.} \ref{fig:4a}, \ref{fig:4c}). \color{black} The averaging time is 15 times smaller than the precession timescale $\tau_{p}=2\pi/\omega_p$.}
	\end{figure}

We shall now explore the evolution of the patterns from the onset of convection
 into the supercritical regime, focusing on two values of the Ekman number 
$E=1.15\times 10^{-5}$ and $E=6.36\times 10^{-6}$, for critical parameters 
 $R_c=(Ra-Ra_c)/Ra_c$ in the range $0.1$--$13$.\\
 Figures \ref{fig:4a} to \ref{fig:4f} illustrate the 
development of convective patterns for a criticality in the range  
$0.1 \le R_c < 12$ and $E=1.15\times 10^{-5}$. 
Near the onset of convection ($R_c=0.13$, figure \ref{fig:4a}) 
we observe a 
behaviour found in the linear stability analysis performed by \cite{goldstein1993convection} on a rapidly rotating 
cylinder with non-slip boundary conditions at the top and bottom and adiabatic 
boundary conditions on the side walls. In this study, the authors characterised
 two different types of convective mode at onset, labelled fast and slow. The 
fast modes correspond to motion at the edge of the cylinder. The slow modes 
describe convective patterns localised at the centre of the cylinder. 
\cite{goldstein1993convection} also showed that the changeover between fast 
and slow modes is strongly dependent on the aspect ratio of the cylinder. On 
figure \ref{fig:4a}, we observe a structure similar to a fast mode of 
azimuthal wavenumber $m=2$. For this wavenumber,
\cite{goldstein1993convection} predicted that the fast
mode was the most unstable below an aspect ratio of 
$1.84$, a condition satisfied in the TC geometry of our experiment. \color{black}Note, however that 
the fast/wall mode structure is not as sharp as for higher levels of criticality. This is 
in part due to the technical difficulty is accurately resolving the small velocity differences 
near the onset. It may also reflect that the TC boundary exerts a somewhat weaker influence 
near the onset than the solid walls of \cite{goldstein1993convection}'s cylinder.\color{black}
\\
For \color{black}$R_c=1.26$ (Fig. \ref{fig:4b}), \color{black}several columns are gathered around 
the centre of the TC. These are reminiscent of structures which 
\cite{aurnou2003experiments} call quasi-geostrophic modes, with the 
difference that in their experiment, dye visualisations suggested that
 they were forming on the outside of the TC. 
 When further increasing $R_c$ ($R_c=4.58$; figure \ref{fig:4c}),
 we see an interaction between these centre modes
 and "wall modes" forming near the side boundary. 
On figures \ref{fig:4d} and \ref{fig:4e}, we observe that for
$R_c \ge 6.29$ the centre modes are merging into one larger structure, which 
evolves into a large central, retrograde vortex in the last stage of the 
convection observed in the experiment (figure \ref{fig:4f}, $R_c=11.27$).
The evolution towards a central retrograde vortex for critical parameters 
exceeding $R_c=10$ was also found by \cite{zhong1993rotating} in experiments 
on convection in a rotating cylinder of aspect ratio $\Gamma=1$ (\emph{i.e.} wider 
than in the present case).\\ 

\begin{figure}
	\subfloat[$R_{c} =  0.35$, \color{black} Localised structure near the centre at onset.]{\includegraphics[width = 0.5\textwidth]{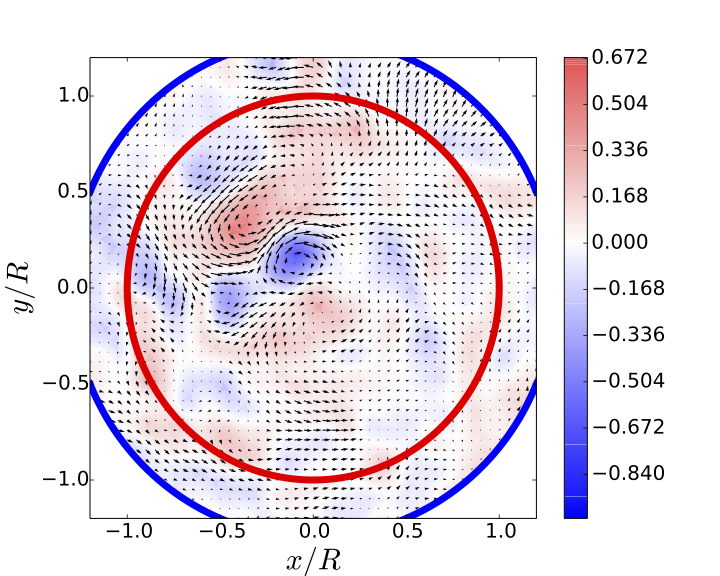}\label{fig:5a}} 
	\subfloat[$R_{c} =  1.26$, \color{black}modulated modes.]{\includegraphics[width = 0.5\textwidth]{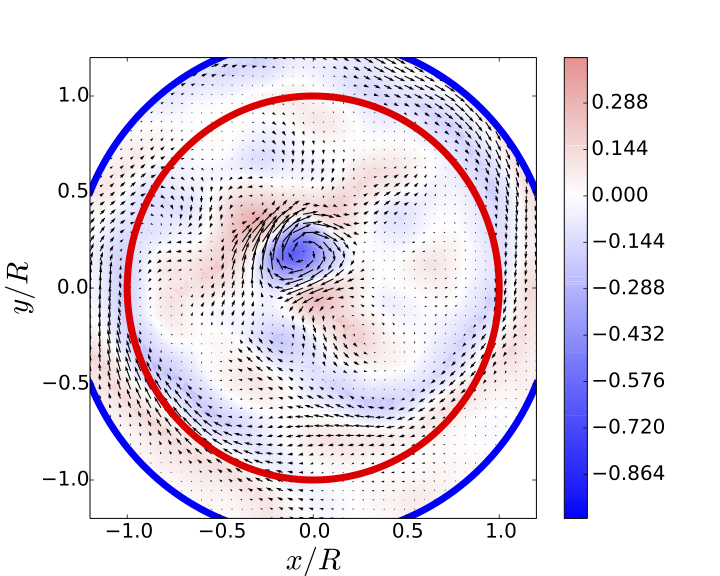}\label{fig:5b}} \\
	\subfloat[$R_{c} =  2.32$, \color{black}wall modes interaction.]{\includegraphics[width = 0.5\textwidth]{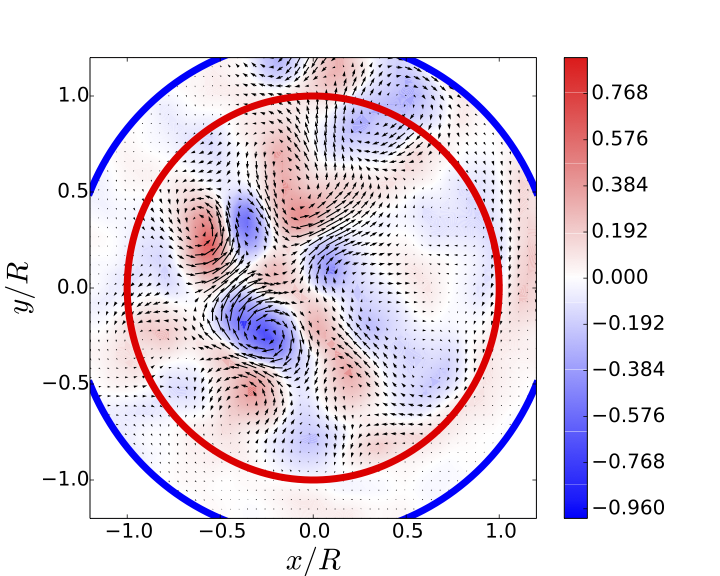}\label{fig:5c}} 
	\subfloat[$R_{c} =  4.40$, \color{black}merging of centre modes.]{\includegraphics[width = 0.5\textwidth]{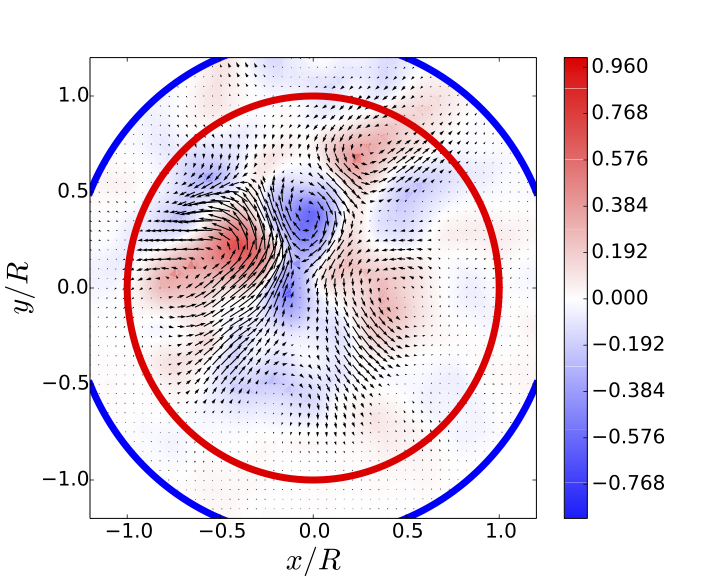}\label{fig:5d}} \\
	\subfloat[$R_{c} =  8.17$, \color{black}merging of centre modes.]{\includegraphics[width = 0.5\textwidth]{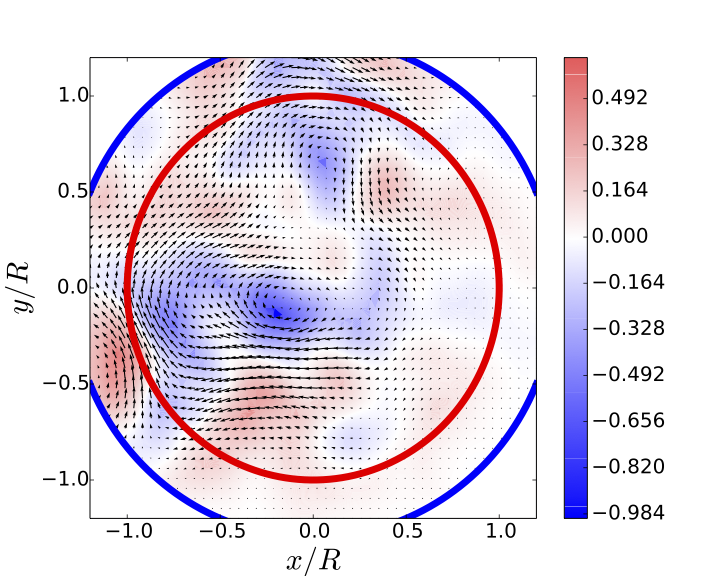}\label{fig:5e}} 
	\subfloat[$R_{c} =  11.91$, \color{black}central retrograde vortex.]{\includegraphics[width = 0.5\textwidth]{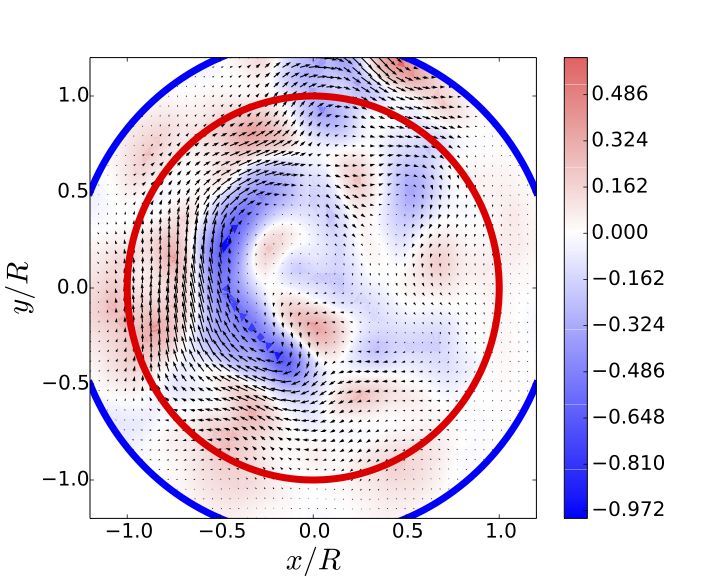}\label{fig:5f}} \\
	\caption{\label{fig:hor2}Time averaged velocity fields (arrows)
          with vorticity field (colorbar) in the horizontal plane at $z/d=3/4$ 
for different
          Rayleigh numbers with $E=6.36\times 10^{-6}$. The blue line represents
          the boundary of the glass dome. The red line represents the position
          of the heater that defines the TC. \color{black} Wall modes can be identified where vorticity 
extrema are present near the boundary of the TC (\emph{e.g.} \ref{fig:5c}). \color{black} The averaging time is 15 times
          smaller than the precession
timescale $\tau_{p}=2\pi/\omega_p$.}
	\end{figure}

Figures \ref{fig:5a} to \ref{fig:5f} show the flow at a smaller Ekman number, 
$E=6.36\times 10^{-6}$, for $R_c \in [0.35\times Ra_{c},11.91\times Ra_{c}]$. 
Near onset ($R_c=0.35$, figure \ref{fig:5a}), 
the convective patterns are much smaller than for a comparable criticality at 
$E=1.15\times 10^{-5}$. When the criticality increases, we observe modes that 
correspond to the modulated modes described by \cite{goldstein1993convection}. 
These feature spiralling arms (Fig. \ref{fig:5b}) and were also observed 
experimentally by \cite{zhong1993rotating} 
for a comparable level of criticality ($R_c = 2.56$). \cite{zhong1993rotating} 
showed that these spiralling patterns arise from the outer wall of the cylinder 
as azimuthal mode. In the TC, we observe such an azimuthal mode with a 
corresponding wavenumber $m=2$. 
Such large differences between supercritical patterns at different Ekman numbers
 were theoretically predicted by \cite{goldstein1993convection}, whose analysis 
shows that when $E$ is varied, the lowest critical Rayleigh number is 
\color{black}alternatively \color{black}achieved by either a fast or a slow precessing 
mode. At higher criticality, convective patterns reflect a combination of wall modes 
and centre modes (at 
$R_c=2.32,4.40,9.17$ on figures \ref{fig:5c}, \ref{fig:5d} and \ref{fig:5e}), as
 for $E=1.15\times 10^{-5}$. Similarly to the more slowly rotating case, 
convective patterns converge towards a central retrograde structure at the 
highest levels of supercriticality explored here (Fig. \ref{fig:5f}).\\

For both values of $E$, the main features of the flow patterns (alternative presence of fast 
and slow modes near the onset of convection depending on $E$, evolution towards a large 
retrograde vortex for $R_c\gtrsim10$) support the view that the 
convection in the TC behaves as convection in a solid cylinder 
rather than in an infinite plane layer. It is also interesting to note that 
the flow patterns discovered by \cite{goldstein1993convection} and 
\cite{zhong1993rotating} at relatively high values of $E$ ($\sim 10^{-3}-10^{-2}$) remain dominant at 
the much lower values explored in our experiment ($\sim 10^{-6}-10^{-5}$). Such robustness may indicate 
that convection is in an asymptotic regime of rotation as far as flow 
structures are concerned and that similar structures might also be found in 
regimes of even more rapid rotation, \color{black}such as the Earth's.\\
\color{black} It should also be mentioned that other authors mention the existence of instabilities 
at the rim of the TC (see \cite{cui2001_jfm,aurnou2003experiments}). While it is difficult to distinguish such 
instabilities from convective instabilities within the cylinder form visualisations only,
\cite{maxworthy1994_jpo} found that such instabilities occur for a Rossby number based on 
vertical motion $Ro^*=u_z/2 \Omega$ greater than 0.28. This parameter remains one or two orders of magnitude below this value in most of our experiments, and below 0.1 in the most supercritical cases (for $R_c\simeq 10$). This is a good indication that the structures we observe are convective patterns, rather than rim instabilities.\color{black}
\subsection{Evolution of the plume size and precession 
frequencies in supercritical regimes \label{sec:evol_super}}
\begin{figure}
  \centerline{\includegraphics[width=1\textwidth]{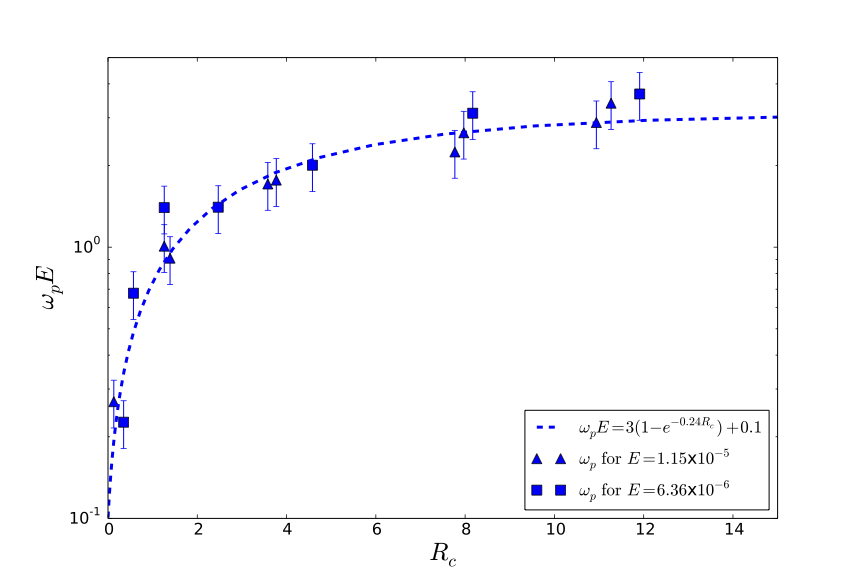}}
  \centerline{\includegraphics[width=1\textwidth]{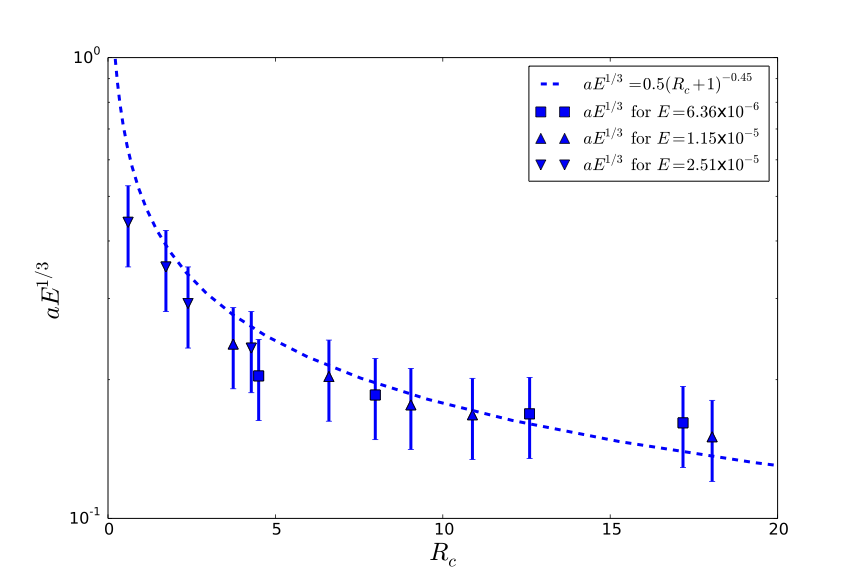}}
  \caption{\label{fig:wave} Evolution of the precession $\omega_{p}$
    and the wave number $a$ as function of the degree of criticality.
 \color{black} Raw and interpolation data for this figure are available as the supplementary material.
}
\end{figure}

The succession of supercritical patterns is reflected in the evolution of 
the dominating wavelengths and precession frequencies with criticality $R_c$ 
gathered on Fig. \ref{fig:wave}.
We find that beyond the onset of convection the dominating wavenumber follows a scaling 
of the form
\begin{equation}
aE^{1/3}=(0.5\pm0.025)\times(R_{c}+1)^{-0.45\pm 0.05}.
\end{equation}
\color{black}
This scaling extends the scaling for $a_{c}(E)$ found at the onset
of convection to supercritical regimes. It implies that as convection becomes 
more intense, the flow rearranges itself with fewer larger 
structures that are more efficient to carry the heat flux
across the TC up to the point where only 
one structure is left for $R_c \gtrsim 10$.

The precession frequency  first sharply increases in the weakly supercritical regime 
and subsequently saturates. It is nevertheless difficult to tell whether an asymptotic 
value is reached in the limit $R_c\rightarrow\infty$. Using a Landau model for the 
bifurcation,  \cite{goldstein1993convection} showed that in weakly supercritical 
regime, the precession normalised by $\Omega$ should vary as: 
\begin{equation}
\omega_{p}=E^{-1}(\delta-\phi R_{c}) + O(R_{c}^{2}),
\label{eq:omega_wsupercrit}
\end{equation}
where the values of constants $\delta$ and $\phi$ depend on the aspect ratio 
of the cylinder \color{black} (with $\delta\rightarrow0$ in the limit of large aspect ratio
to recover the stationary onset of convection in an infinitely extended 
plane layer, see \cite{chandrasekhar1961hydrodynamic}). \color{black}
The values of $\delta=0.1$ and $\phi=5$ were obtained experimentally
 by \cite{zhong1993rotating} and \cite{ecke1992hopf} for a cylinder of aspect 
ratio $\Gamma=1$ and values of $E$ greater than $10^{-3}$.
The variations of $\omega_p(r_c)$ extracted from experiments at several Ekman 
numbers for $R_c>0$ are reported in Fig. \ref{fig:wave}. They are best fitted 
over the widest measured range of values by an exponential law of the form:
\begin{equation}
\omega_{p}=(3\pm0.5)\times E^{-1}(1-e^{(-0.24\pm0.01)R_{c}}) + 0.1 \times E^{-1},
\label{eq:omega_supercrit}
\end{equation}
Expanding (\ref{eq:omega_supercrit}) to $\mathcal O(R_c)$ and identifying with 
(\ref{eq:omega_wsupercrit}), we find that $\phi=0.534\pm0.105$.
The value of $\delta$ is consistent with the findings of \cite{ecke1992hopf}, 
but the precession frequency appears to vary significantly less in the 
supercritical regime in the present case 
than in the case of a solid rotating cylinder. It is, however, 
difficult to attribute \color{black} this difference to any of the factors that \color{black} differ 
between the two problems: the difference in shape of the upper domain boundary, 
a different aspect ratio, a three-order of magnitude difference in 
the Ekman number, different mechanical and thermal boundary conditions at the 
lateral boundary of the cylinder. Nevertheless, the precessing motion 
at onset and in supercritical regimes adds further support to the view that 
the phenomenology of convection in the tangent cylinder obeys the same 
mechanisms as in a solid rotating cylinder, both at the onset and in 
 supercritical regimes.
\section{Heat flux and thermal wind \label{sec:heat_wind}}
\subsection{Heat Flux through the Tangent Cylinder}
The efficiency of convection is best measured by its ability to transport heat 
through a fluid layer. 
Here we characterise the variations of the heat flux with $Ra$ and 
$E$ by means of the 
Nusselt number which represents the ratio of the heat flux in the fluid to the 
purely conductive heat flux:
\begin{equation}
Nu=\frac{Fd}{k\Delta T},
\end{equation}
where we recall that $F$ is the heat flux through the top surface of the heater, obtained 
from the difference between the temperature of the heat-carrying fluid at the 
inlet and outlet of the heater and $d$ is the height of fluid above the heater.
The variations of 
$Nu$ with $Ra$ are reported in Fig. \ref{fig:heatflux}
for several values of $E$.
\color{black}
Since $F$ is measured at the heater, it reflects the total heat transferred across the entire dome. As all of the flux transits through the TC, $F$ is an upper bound for the total heat flux through the upper surface of the TC because thermal losses occur at the lateral boundary of the heater. Given the low values of the velocity near that boundary, these losses are mostly likely predominantly of conductive nature, especially at low criticality.
 \color{black} Two regions of parameters clearly appear.
In the large Ra limit, the heat transfer becomes independent of the rotation and 
follow a scaling of the form (see top graph):
\begin{equation}
Nu=(0.2 \pm0.04) Ra^{0.33\pm0.03}.
\label{eq:nuscal_hira}
\end{equation}
This suggests that in these strongly supercritical regimes, heat transfer are 
not influenced by the Coriolis force anymore. Indeed, in the range of 
Rayleigh and Prandtl  numbers we considered, similarly low exponents of $Ra$ in the range 
1/4-3/7 are expected for plane non-rotating Rayleigh-Benard convection \citep{grossmann2000_jfm} and 
in spherical shells with radial gravity \citep{gastine2015_jfm}.  
At moderate Rayleigh numbers, by contrast, ($Ra\gtrsim8\times10^8$ for $E=1.15\times10^{-5}$, 
$Ra\gtrsim1\times10^9$ for $E=6.35\times10^{-6}$  and 
$Ra\gtrsim1.5\times10^9$ for $E=4.46\times10^{-6}$, see bottom graph), 
$Nu$ exhibits a strong dependence on $E$ as well as $Ra$, of the form:
\begin{equation}
Nu=(0.38\pm0.02)\times Ra^{1.58\pm0.06}E^{2\pm0.04}.
\label{eq:nuscal_lowra}
\end{equation}
This law is very close to the theoretical scaling of $Nu\sim 0.15 Ra^{3/2}E^2$ put forward 
by \color{black} \cite{julien2012_prl} (see also \cite{gastine2016_jfm}) \color{black}as a 
signature of the diffusivity free-regime. 
These authors found that this scaling was verified in rotating convection in a spherical shell
with radial gravity for a range of values of $Ra$ and $E$ that is essentially the same 
as its range of validity in LEE. Their theoretical scaling argument, however, does not invoke 
geometry nor the orientation of the buoyancy force. It relies on the assumptions that (i) the 
largest contribution or the temperature gradient originates in the bulk and not in boundary 
layers, (ii) in the limit of no rotation, the ultimate regime of classical Rayleigh-Benard 
convection is recovered \citep{kraichnan1962_pf}, (iii) in the limit where rotation dominates inertia, the heat flux solely depends on $R_c$ \cite{king2012_jfm}. 
Hence, this scaling can be reasonably interpreted as the signature of the diffusivity-free 
regime of rotationally-dominant convection in the TC configuration too. 
\color{black} 
Experiments and numerical simulations on convection in a rotating cylinder filled with water 
\citep{cheng2015_gji, king2012_jfm} also exhibit a 
similar transition between steep variations of $Nu(Ra)$ in the weakly supercritical regime and 
a more a strongly supercritical regime where $Nu\sim Ra^{0.32}$ that is very close to 
(\ref{eq:nuscal_hira}). On the other hand, the scaling law at low $Ra$, was found to strongly depend 
on $E$, with an exponent varying from $Nu\sim(RaE^{4/3})^{6/5}$ at low Ekman numbers to 
$Nu\sim(RaE^{4/3})^{3.6}$ (for $E=10^{-7}$).
\color{black}

\color{black}
\begin{figure}
  \centerline{\includegraphics[width=0.8\textwidth]{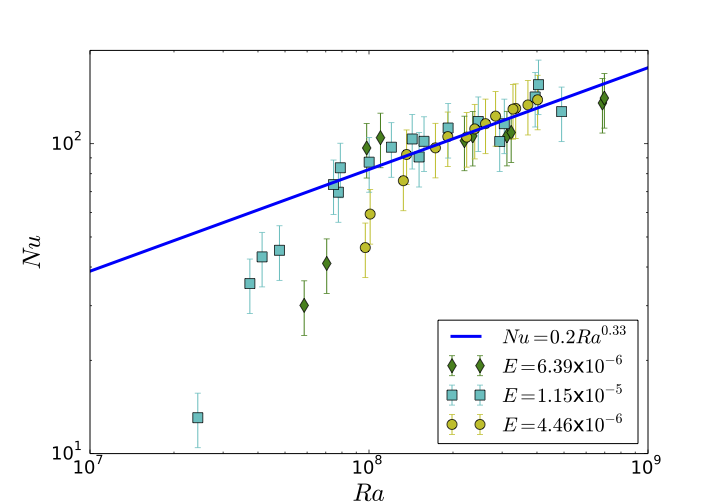}}

  \centerline{\includegraphics[width=0.8\textwidth]{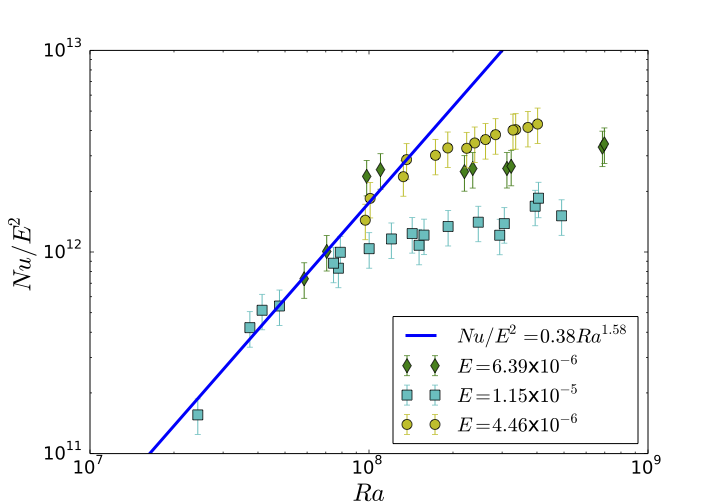}}
  \caption{\label{fig:heatflux} \color{black}Variations of the Nusselt number $N$ emphasising  
the scaling $Nu/E^{2}\simeq 0.2 Ra^{0.33}$ for large Rayleigh numbers (top) and the 
the scaling $Nu\simeq 0.38 E^2 Ra^{1.58}$ at moderate Rayleigh numbers (bottom). 
Raw and interpolation data for this figure are available as the supplementary material.}
\end{figure}

An alternative but equivalent way to analyse the variations of the heat flux 
consists of seeking how flux normalised by rotation varies with the Rayleigh number based on the 
heat flux itself. One advantage of this approach is to quantify convection 
in terms of the available buoyancy. Buoyancy may indeed result from a heat flux 
at the boundary but also from a solute mass flux, as in the core of the Earth 
and other planets. On this basis, we follow 
\cite{aubert2001systematic, aubert2005steady, christensen2006scaling} and 
introduce the modified Nusselt number, Rayleigh number and a modified 
diffusionless Rayleigh number based on the heat flux, respectively defined as

\begin{equation}
Nu^{*}=Nu\times E \times Pr^{-1},
\label{eq:nustar}
\end{equation}  

\begin{equation}
Ra^{*}=Ra\times E^{2} \times Pr^{-1},
\end{equation}

\begin{equation}
Ra_{q}^{*}=Ra^{*}Nu^{*}.
\label{eq:raq}
\end{equation}  

\color{black}
\cite{cheng2016_espl} stresses that scalings of the form $Nu\sim Ra^\alpha$ 
and $Nu^*\sim {Ra_q^*}^\beta$ satisfied $\beta=\alpha/(1+\alpha)$. Here 
$\alpha=0.33\pm0.03$, $\alpha/(1+\alpha)\in[0.23,0.27]$ and $\beta=0.3$. 
\color{black} Hence, these scalings offer an alternative way of representing 
the data from figure \ref{fig:heatflux}, rather than new data.    

\color{black}
\cite{christensen2002zonal} found that these quantities obeyed a scaling law 
of the form 

$Nu^{*} \sim (Ra_{q}^{*})^{5/9}$. 

This result was obtained with numerical simulations in a spherical shell 
geometry for $Ra_{q}^*\in[10^{-7},10^{-3}]$. \cite{aurnou2007planetary} 
identified this scaling as indicative of a rapidly rotating regime. 
Further, using the data of \cite{sumita2003experiments},
it was shown that for low values of $Ra_{q}^{*}$ the relation between $Nu^{*}$ 
and $Ra_{q}^*$ was better fitted with a power law of the form $Nu^{*} \sim (Ra_{q}^*)^{0.29}$.
In the present configuration, $Ra_{q}^*$ varies between
 $10^{-10}$ and $10^{-6}$ and therefore falls within a similar
range to the experiments of \cite{sumita2003experiments}.
Collapsed data reported in Fig. \ref{fig:heatflux}
\color{black} show that points in the regime following the $Nu\sim Ra^{0.33}$ law
obey a scaling close to that found by these authors:

\color{black}
 \begin{equation}
 Nu^{*} = (0.0046\pm0.0005) \times (Ra_{q}^{*1\pm0.03})^{0.26\pm0.04}. 
\label{eq:nustar}
  \end{equation} 

\color{black}The exponent being closer to 0.29 than 5/9 confirms that this regime
is one the flow is outside the quasi-geostrophic regime, where convection-driven 
inertia plays an important role. 
\color{black}

\begin{figure}
  \centerline{\includegraphics[width=0.8\textwidth]{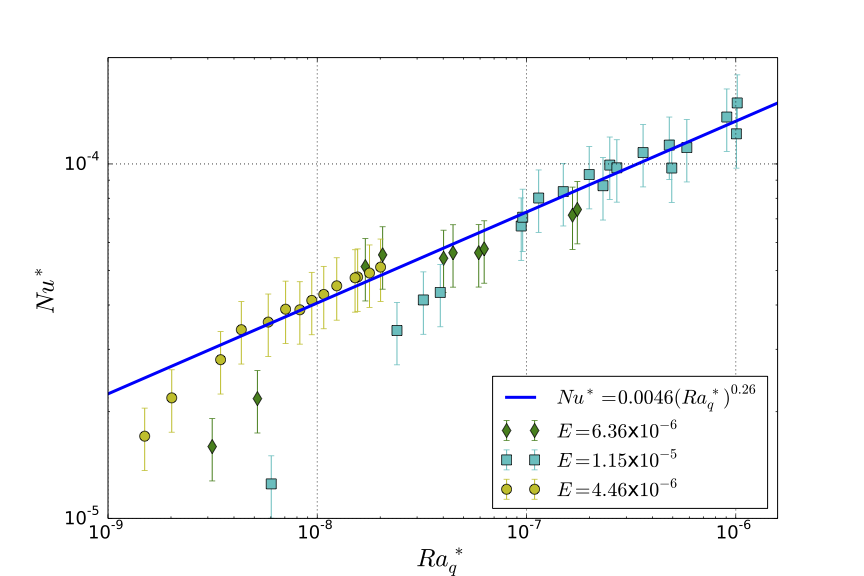}}
  \caption{\label{fig:heatflux} Variations of $Nu^{*}$ with $Ra_{q}^{*}$. 
\color{black}Raw and interpolation data for this figure are available as the supplementary material.}
\end{figure}

\subsection{Thermal wind}
In axisymmetric geometry, the balance between buoyancy
and Coriolis forces gives rise to azimuthal motion,
seen through the balance between the curl of these forces:

\begin{equation}
  \frac{\partial u_\theta}{\partial z}\sim-\frac{g\beta}{2\Omega}
  \frac{\partial T^\prime}{\partial r},
\label{eq:tw}
\end{equation}

where $\beta$ is the coefficient of thermal expansion and $T^\prime$ is the
temperature perturbation. The strongest retrograde motion 
is found at the higher
latitude, such as those of our PIV planes represented in figures 
\ref{fig:hor1} and \ref{fig:hor2}, located at a latitude of 
$51.5^{\circ}$ (or, equivalently at $z/d=0.75$ above the heater). 

On the other hand, the direction of the thermal wind may reverse at low 
latitude. Hence, we extract from these measurements a radial profile of the 
azimuthal wind at two different latitudes (see figure \ref{fig:general}) by 
means of an azimuthal and time average of the azimuthal velocity
$\langle u_\theta(r,z) \rangle_{\theta t}$: additionally to the high  
latitude, 

the second 
latitude we consider corresponds to a plane close to the heater surface 
($20^{\circ}$, or, equivalently at $z/d=0.167$ above the heater).

\begin{figure}
\subfloat[]{\includegraphics[width = 0.5\textwidth]{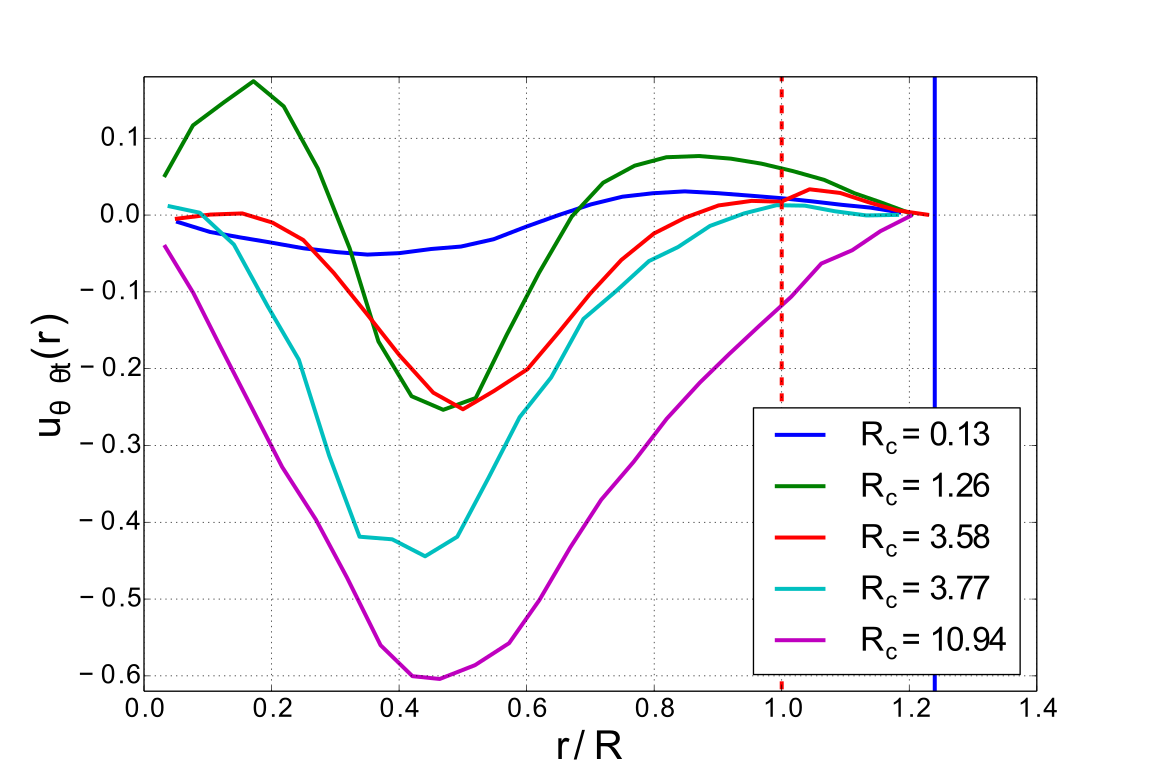}\label{fig:tw_a}} 
\subfloat[]{\includegraphics[width = 0.5\textwidth]{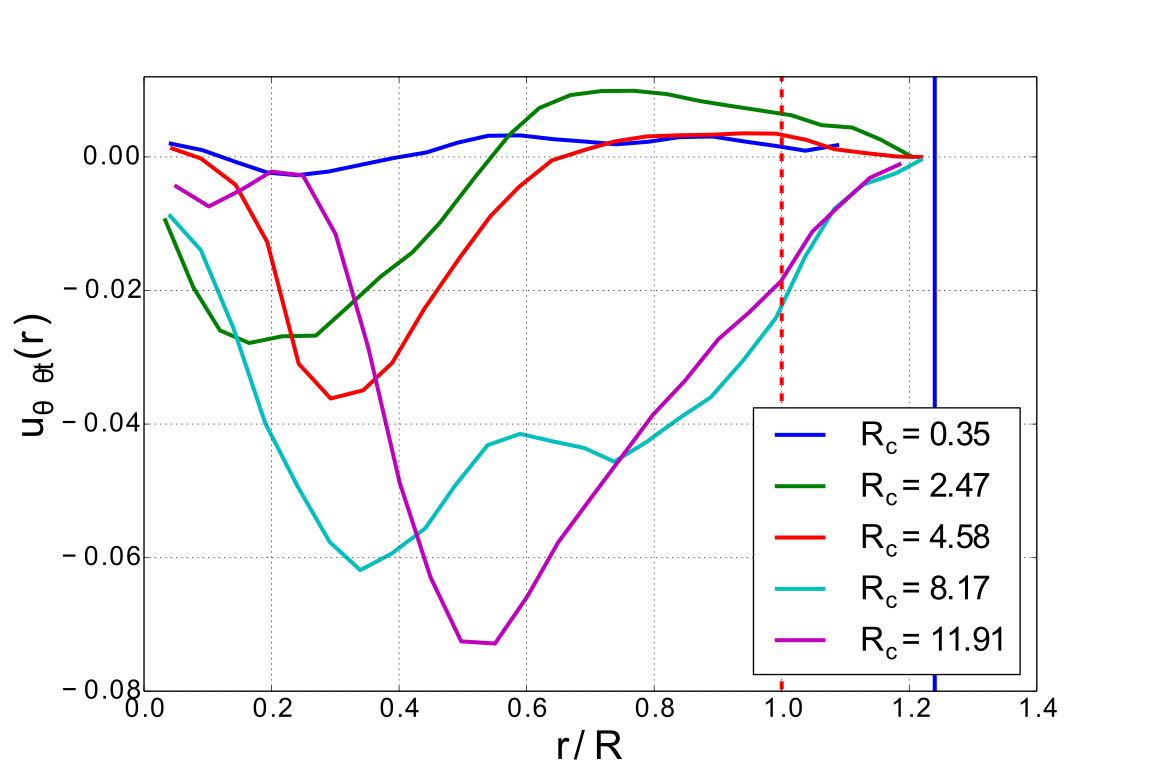}\label{fig:tw_b}}\\
\subfloat[]{\includegraphics[width = 0.5\textwidth]{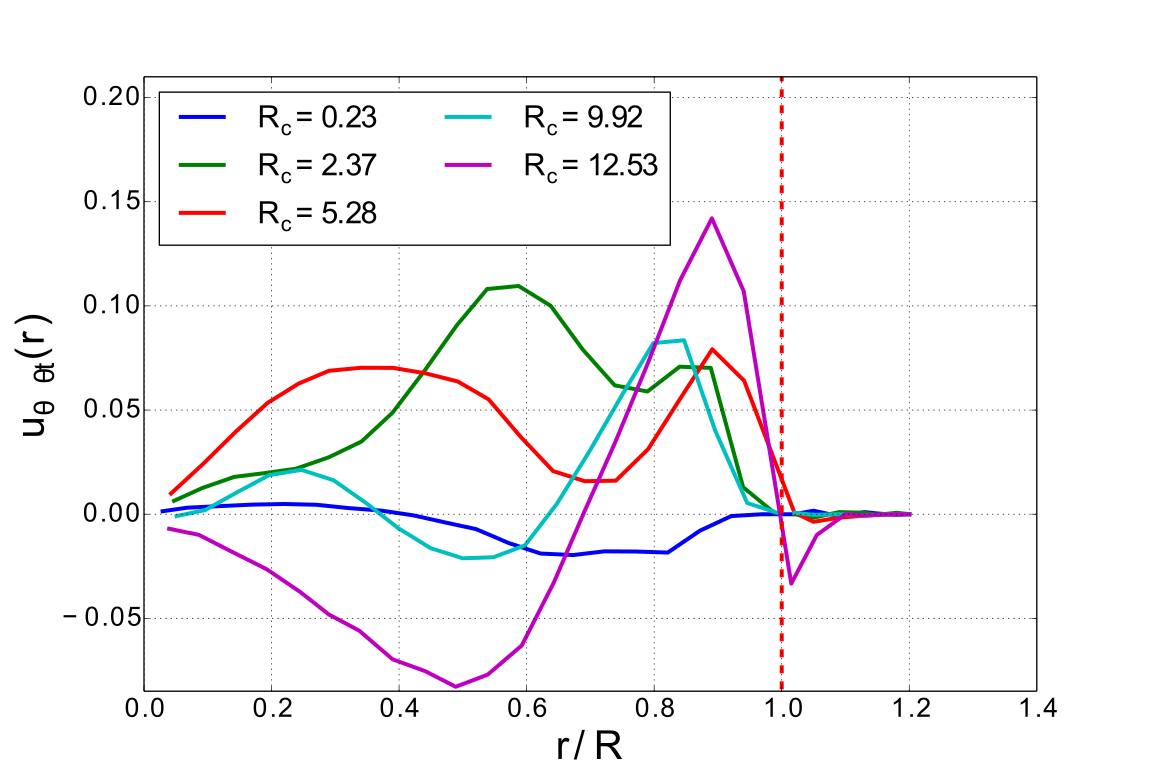}\label{fig:tw_c}} 
\subfloat[]{\includegraphics[width = 0.5\textwidth]{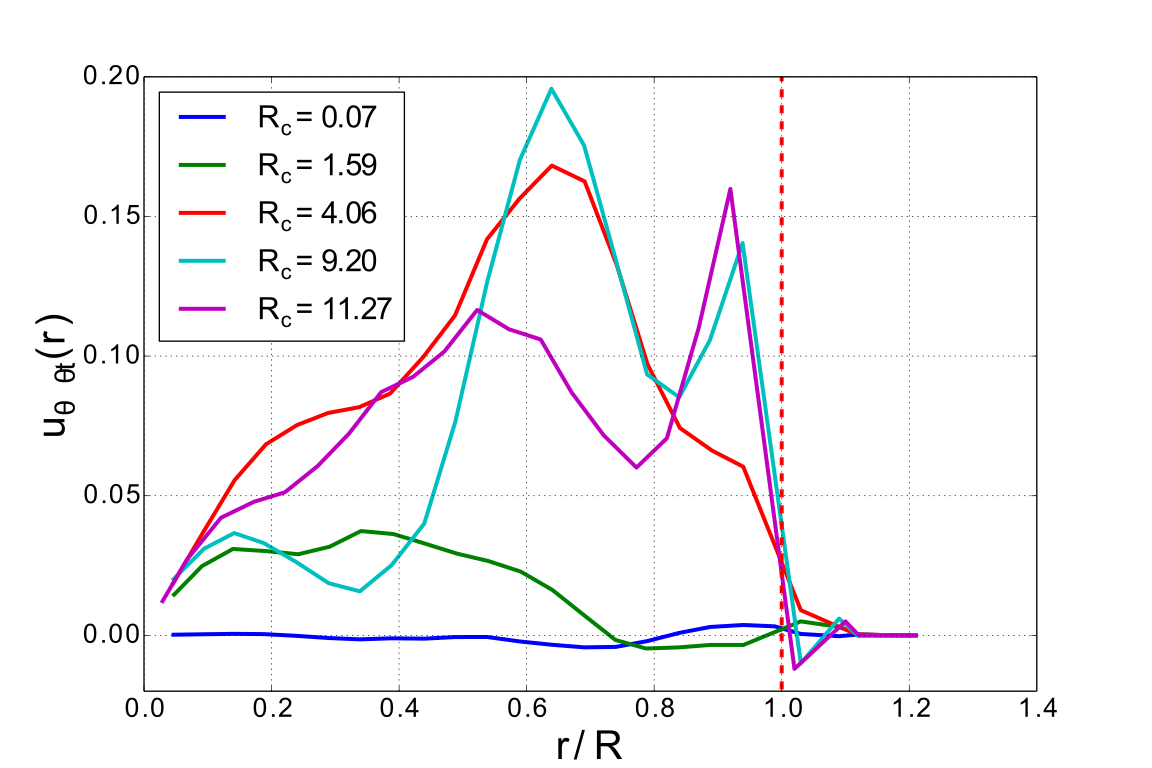}\label{fig:tw_d}}\\
\caption{Azimuthally and time-averaged radial profiles of azimuthal velocities $\langle u_{\theta}\rangle_{\theta t}(r)$, for several levels of criticality. Top: data from horizontal plane  at $51.5^{\circ}$. Bottom, data from horizontal plane at $20^{\circ}$. Left $E=1.15\times 10^{-5}$, Right: $E=6.36\times 10^{-6}$.
Vertical dashed line: TC boundary, vertical solid line: position of the glass dome.}
\end{figure}

Results are plotted on figures \ref{fig:tw_a} to 

\ref{fig:tw_d}.\\ 
At high latitude (figures \ref{fig:tw_a} and \ref{fig:tw_b}), the profiles 
exhibit a strong negative maximum followed by slightly positive 
values at larger radii, corresponding to a strong retrograde motion surrounded 
by a slightly prograde motion. Near the onset of convection, some prograde 
motion exists near the centre, that disappears as the flow becomes more 
supercritical. Despite the succession of different patterns observed in the 
supercritical regime, the
intensity of the retrograde motion steadily increases with criticality.
For strongly supercritical flows ($R_c\gtrsim11$), velocity patterns show a 
single structure slightly outgrowing the TC
(we shall see in section \ref{sec:confinement} that motion outside the TC does 
not necessarily imply that the Taylor-Proudman constraint is broken). This 
phenomenon appears for 
slightly higher criticality at higher values of
$E$ ($R_c\gtrsim9$ at $E=1.15\times10^{-5}$ and $R_c\gtrsim11$
at $E=6.36\times10^{-6}$). In both cases, 
this confirms that in the most supercritical regimes, retrograde 
motion progressively invades the high-latitude region of the TC, where 
the flow becomes dominated by a large central retrograde vortex.\\
Profiles at low latitude, nearer the solid inner core provide a better picture 
of the three-dimensional structure of the thermal wind (Figures \ref{fig:tw_c} 
and \ref{fig:tw_d}). In both cases, a prograde wind first develops near the 
solid core at low levels of criticality and increases in intensity with $R_c$. 
At $E=1.15\times10^{-5}$, the prograde wind starts weakening at the centre 
from $R_c=2.37$ and starts becoming retrograde around $R_c=9.92$.
At higher rotation, weakening of the prograde wind near the solid core occurs 
only for $R_c>9$ and we did not reach a regime where it reversed. 
This difference in levels of criticality for the weakening and reversal of 
the thermal wind near the solid core can be understood from the scaling for the 
critical Rayleigh number \color{black} $Ra_c\sim E^{-4/3}$ \color{black} and (\ref{eq:tw}) which imply that 
$\partial_z u_\theta\propto (R_c+1)E^{1/3}$. From this scaling, thermal wind 
with a given vertical gradient of $u_\theta$ is expected to occur at increasingly high levels of criticality when $E$ is decreased. Hence, the larger velocity 
gradients corresponding to a thermal wind flowing in opposite directions at high and low latitudes survive at higher levels of criticality when $E$ decreases.\\ 
Interestingly, a clear azimuthal prograde jet is present at the edge of the 
tangent cylinder on the inside. The jet is thinner at faster rotation.
It appears for $R_c\sim 2.37$ at $E=1.15\times10^{-5}$ and 
$R_c\sim 4.06$ at $E=6.36\times10^{-6}$, and its intensity slightly increases 

with criticality.
One cannot but notice the similarity between this structure and the 
non-axisymmetric azimuthal jets near the TC inferred by 
\cite{livermore2017_natgeo} from satellite measurements of 
time variation of the Earth's magnetic 
field. Although the forces driving the 
jet remain unknown, the authors suggest that they are most likely of magnetic 
origin, an effect that is absent in the present study. Azimuthal jets were also noticed 
in measurements in the flow patterns inferred from measurements of the Earth magnetic field \citep{hulot2002small}.\\

\begin{figure}
  \centerline{\includegraphics[width=0.8\textwidth]{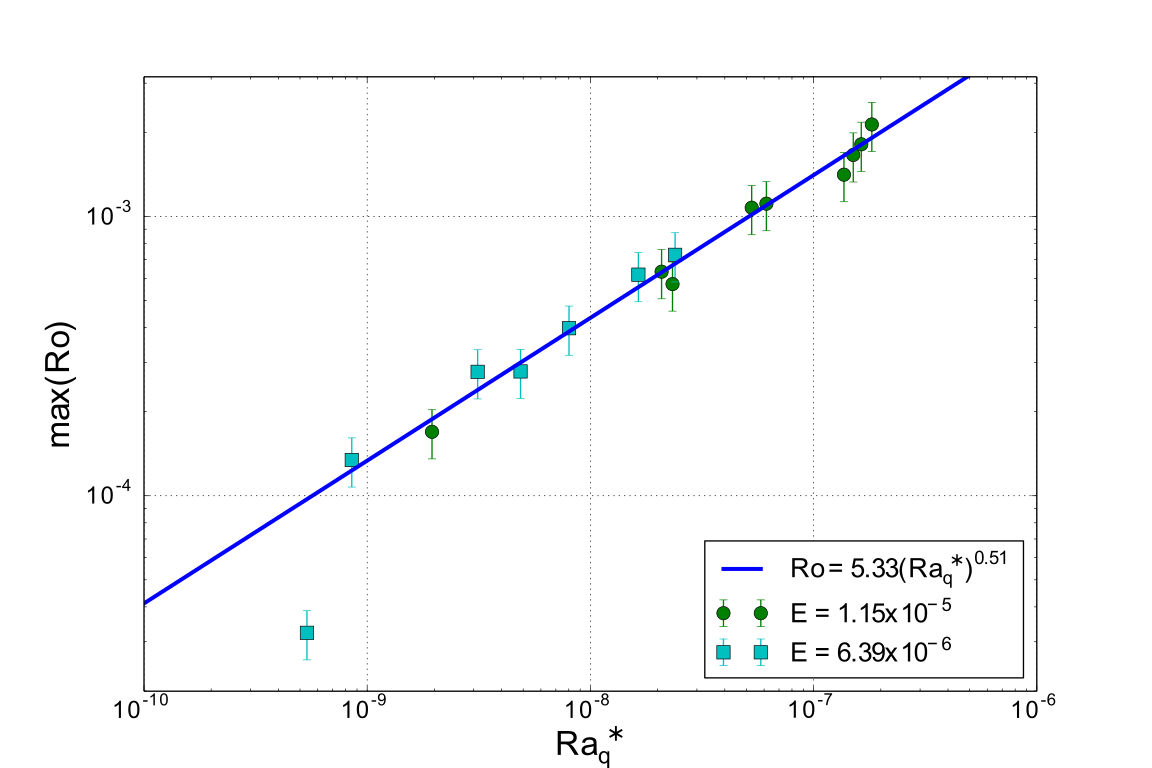}}
  \caption{\label{fig:rossby} Variations of the Rossby number Scaling $Ro$ with $Ra_{q}^*$. 
 \color{black} Raw and interpolation data for this figure are available as the supplementary material.
}
\end{figure}

To conclude the characterisation of the thermal wind, we shall quantify its 
intensity in terms of the Rossby number $Ro$, and its variations with the 
flux-based Rayleigh number, $Ra_{q}^{*}$. The Rossby number, defined as

\begin{equation}
Ro=\frac{U_m}{2R\Omega}
\end{equation}  

measures the ratio of inertia to the Coriolis force, where $U_m$ is chosen 
as the maximum retrograde velocity in the averaged profiles of the thermal 
wind measured in the high latitude plane (reported in figures \ref{fig:tw_a} 
and  \ref{fig:tw_b}).
Our results, reported in figure \ref{fig:rossby}, obey the scaling \color{black} 

\begin{equation}
 Ro=(5.33\pm0.3) \times (Ra_{q}^*)^{0.51\pm0.04}. 
\label{eq:ro_scaling}
 \end{equation}

This scaling is very close to the scaling relating the azimuthal velocity scale 
$U_\theta$ and buoyancy $B$,  $u_\theta\sim (B/\Omega)^{1/2}$ (here, $Ro\sim {Ra_q^*}^{1/2}$)
 first derived theoretically by \cite{maxworthy1994_jpo} for the 
thermal wind in oceanic convective plumes. \color{black} It is derived from the azimuthal 
curl of a local balance between  Coriolis force and buoyancy
outside the geostrophic regime, where the excess buoyancy and the Rossby deformation radius
are set by the balance between Coriolis and buoyancy forces \citep{maxworthy1994_jpo}. 
\color{black} Both \cite{aurnou2003experiments} and 
\cite{aubert2005steady} report the same scaling when thermal wind is present 
respectively in experiments in a TC geometry and in numerical 
simulations in full spherical shell. Hence (\ref{eq:ro_scaling}) provides further 
evidence that the azimuthal motion we are detecting is thermal wind. \color{black} the vertical 
gradients of azimuthal velocity induce a departure to geostrophy or order $\mathcal O(Ro)$, which 
is significantly smaller than the vertical velocities associated with the plumes. Hence the thermal 
wind may not be the main source of the ageostrophy suggested by the variations of heat flux.
 \color{black}

\section{Effect of the confinement \label{sec:confinement}}
\begin{figure}
\subfloat[]{\includegraphics[width = 0.5\textwidth]{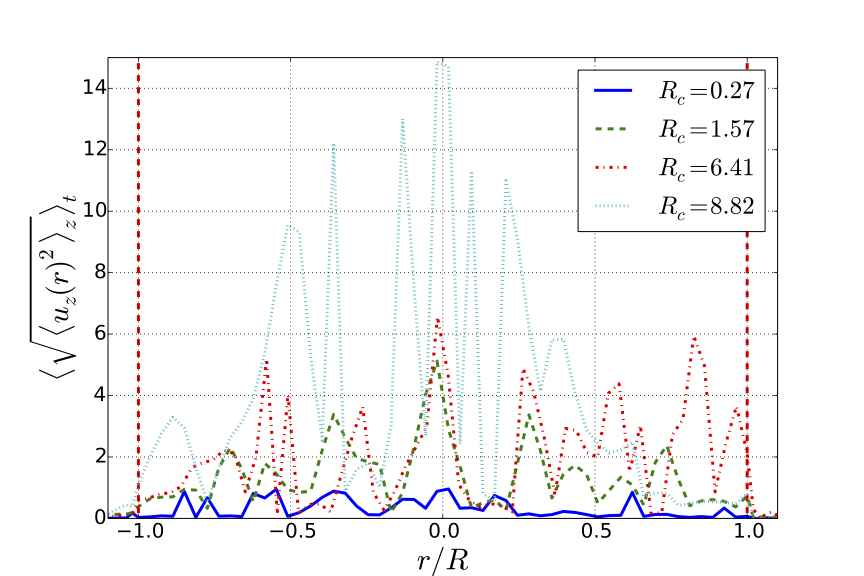}\label{fig:cf_a}} 
\subfloat[]{\includegraphics[width = 0.5\textwidth]{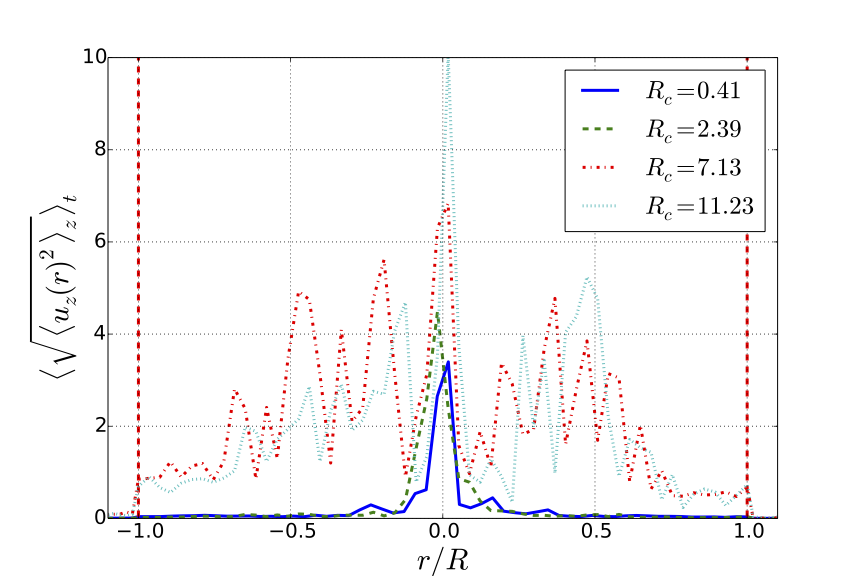}\label{fig:cf_b}}\\
\subfloat[]{\includegraphics[width = 0.5\textwidth]{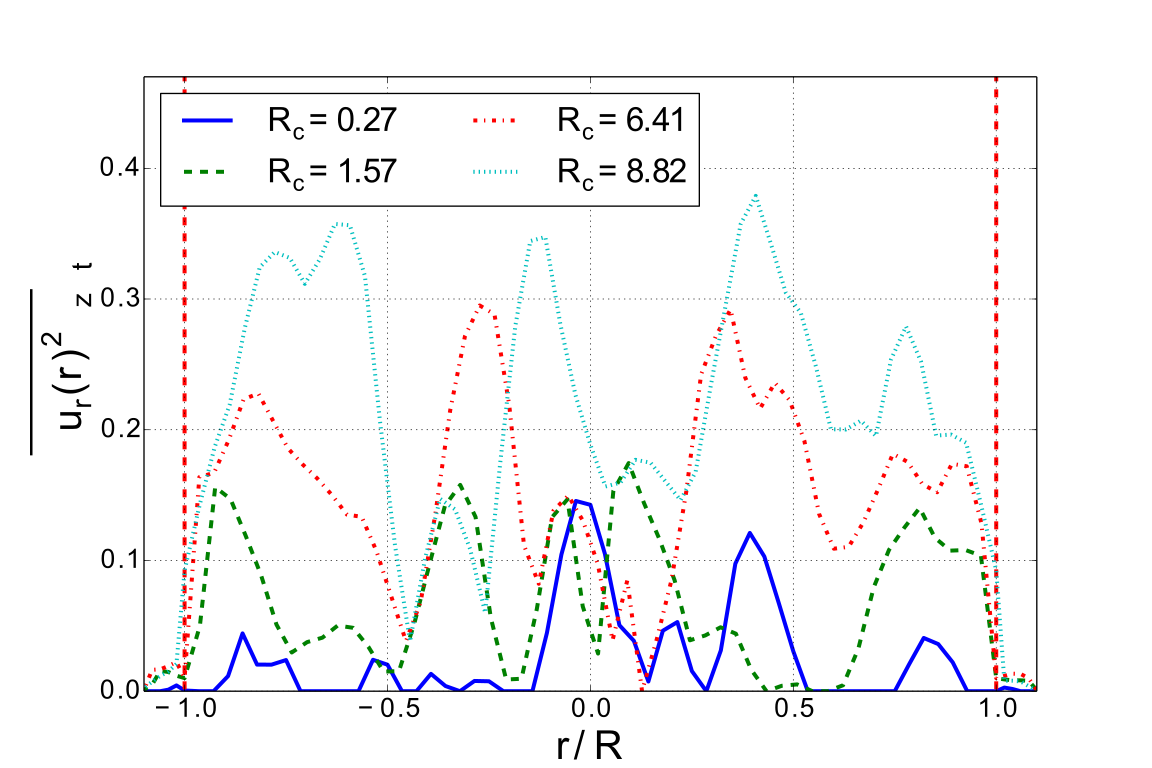}\label{fig:cf_c}} 
\subfloat[]{\includegraphics[width = 0.5\textwidth]{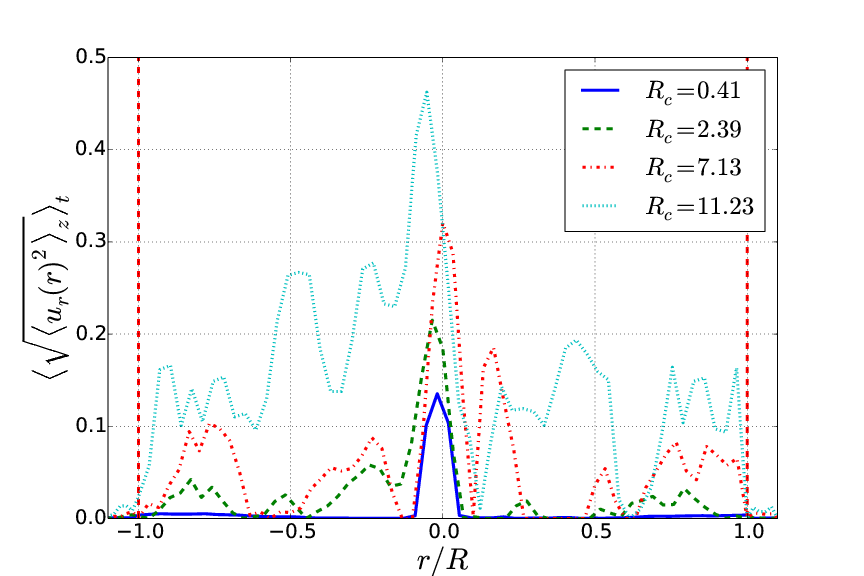}\label{fig:cf_d}}\\
\caption{(a) and (c) $\langle\sqrt{\langle u_{z}(r)^{2}\rangle_{z}}\rangle_t$ (top) and $\langle\sqrt{\langle u_{r}(r)_{r}^{2}\rangle_z}\rangle_t$(bottom) for several increments of supercriticality and $E=1.15\times 10^{-5}$.  (b) and (d) Similarly, $\langle\sqrt{\langle u_{z}(r)\rangle_{z}^{2}}\rangle_t$ and $\langle\sqrt{\langle u_{r}(r)_{r}^{2}\rangle_z}\rangle_t$ for several increments of supercriticality and $E=6.36\times 10^{-6}$. }
\end{figure}

The structure of convection, the scalings for the heat flux 
and the thermal wind present a picture of convection within the TC at low 
Ekman numbers that resembles convection in a cylinder confined by solid walls: 
the steep rise heat transfer at low criticality, where the flow is dominated by rotation
 is followed by a rotation-independent regime at higher criticality.
The similarity with rotating convection in a solid cylinder reflects a prominent role of
background rotation through the TP constraint it imposes on the flow. \color{black} Based on these 
observations, we shall conclude this study by measuring the degree of confinement within the TC. 
Figures \ref{fig:cf_a} and \ref{fig:cf_c} respectively show the $z$-rms 
profiles of vertical and radial velocities for $E=1.15\times 10^{-5}$. 
The radial velocity right across the side boundary of the TC never exceeds a few percent 
of its typical value within the TC so the TP constraint can be seen as 
enforcing a near-impermeable condition there. For low criticality, vertical 
fluid motion is entirely contained within the 
radius of the TC. However, for $R_c\geq6.41$, a slight motion appears 
outside the TC, that remains at approximately the same level as $R_c$ is 
increased. The same is true for the azimuthal wind, which extends slightly 
beyond the TC at moderate to high levels of criticality
(figures \ref{fig:tw_a} and \ref{fig:tw_c}). 
Vertical and azimuthal motions do not, however, directly break 
the TP constraint. Momentum inside the TC is indeed transported across the 
TC boundary by viscous friction. Friction is all the more effective there 
as a free Stewartson layer is expected to develop there, with an 
inner thickness scaling as $E^{-1/3}$ and an outer thickness scaling 
as $E^{-1/4}$ \citep{stewartson1957}.
Such layers are too thin to be reasonably detected 
in our measurements. Nevertheless, the radial profiles of azimuthal velocities 
in figure \ref{fig:tw_a}- \ref{fig:tw_d} show hint of a variation in slope 
across the TC boundary. This local slope is also considerably steeper at 
$E=6.36\times 10^{-6}$ than $E=1.15\times 10^{-5}$. This suggests that
the mechanical condition across the TC boundary is probably closer to 
one of imposed tangential stress of a value determined by the rotation, 
rather than the no-slip condition of a rigid boundary.\\

This tendency is confirmed by the $z$-rms profiles of vertical and radial 
velocities for $E=6.36\times 10^{-6}$. For this faster rotation, the radial 
velocity is found to be exactly zero at the boundary of 
the TC at all levels of criticality, within the precision of our measurements. 
Vertical velocity is also practically zero. The convection even seems  
extinct on the last 20\% of the TC's outer region. Clearly, this behaviour is 
promoted by the fact that the hot boundary of the domain is itself confined 
within the domain. However, the intense convection that ensues would not remain 
confined within the higher latitudes of the TC without a strong influence of 
the TP constraint. Overall, the behaviour is similar to that found at 
$E=1.15\times 10^{-5}$, except that the TP constraint is more strictly 
enforced at equivalent levels of criticality.\\

\color{black}
Note that the curvature of the dome may have an influence on the 
confinement effect. The height under the dome varies by 7.8\% between the centre 
and the edge of the TC. This geometry has 
two consequences: first, the onset of convection is determined by $RaE^{4/3}$, 
which dependents on the local height under the dome $h$ as $h^{1/3}$. This, however, 
implies that 
$RaE^{4/3}$ only varies by 2.5\% across the heater which is unlikely to decide 
where the first convective cells appear. The limited influence of the curvature is 
further confirmed by the fact that the critical Rayleigh number follows practically 
the same scaling as for plane convection. 

\color{black}
Second, the associated TP constraint 
opposes radial motion which slightly reinforces confinement. 
In planetary cores, the spherical shape of the solid core reverses the radial 
variation of height compared to our experiment but it similarly opposes radial 
motion, thus reinforcing 
confinement.  In both cases, however, the confinement associated to the curvature 
of the dome is expected to be of significantly less influence than the 
TP constraint incurred at the outside edge of the TC.

\section{Conclusions and discussion}
The experimental study we conducted was focused on convection in a Tangent 
Cylinder for Ekman numbers in the range $3.36\times10^{-6}$ to 
$4.51\times10^{-5}$, and brought several answers to the four questions set in 
the introduction:\\
First, the critical scalings for the onset of convection in a TC are similar
 to those known for plane convection, albeit with different constants: 
\color{black} the critical Rayleigh number 
$Ra_c=(26\pm4)\simeq E^{-4/3\pm 0.1}$ is \color{black} marginally higher,
 but the critical wavenumber $a_{c}=(0.5\pm0.07)\times E^{-1/3\pm0.05}$ 
is significantly higher than for plane convection. 
Second, this discrepancy has origins in the structure of the critical 
convective plumes, which resemble those found in rotating cylinders, rather 
than the periodic cell pattern of plane rotating convection. As in the former, 
the critical mode is either one of the slow or one of the fast modes identified by 
\cite{goldstein1993convection}, depending on the Ekman number and the aspect 
ratio of the cylinder.
\color{black} In a solid cylinder, the onset of wall modes takes place at a significantly 
lower critical Rayleigh number than the unstable modes of convection in a plane layer. In 
a Tangent Cylinder, however the Taylor-Proudman constraint does not exist in the still base flow, so the confinement that is responsible for the onset of wall modes is absent at such low values 
of $Ra$. Without it, the base flow retains a configuration that is stable up to values of 
$Ra$ for which plane modes are unstable. When these are ignited, however, the TP constraint becomes 
active and favours eigenmodes of the cylindrical geometry. 
 \color{black} Nevertheless, the influence of the virtual TC 
boundary being weaker than that of a solid wall, the plumes at onset retain some 
of the features of that of a plane configuration, with in particular, a size that is 
intermediate between those of plane and cylindrical geometries.
\\ \color{black}
 Third, the loss of translational symmetry in the 
horizontal plane excludes an onset of convection through steady modes, even at 
the moderate values of $Pr$ for which one would expect steady
rather than oscillatory onset in a
 plane configuration. However, the time-dependence takes the form of 
a very slow retrograde precession, instead of waves expected at oscillatory onsets.\\
Fourth, the supercritical regimes exhibit a complex sequence of 
convective patterns, leading to a single, large vortex centred on the cylinder axis when 
the Rayleigh number exceeds approximately 10 times the critical value. At
relatively high latitude, this vortex translates into a coherent, 
retrograde thermal wind. For the Ekman numbers we considered, 
the intensity of the thermal wind, 
measured in terms of the Rossby number obeys a scaling of 
\color{black} $Ro=(5.33\pm0.3)\times (Ra_{q}^*)^{0.51\pm0.04}$  
identical to the scaling found in previous studies 
\citep{aurnou2003experiments} for thermal wind resulting from the 
\color{black} interplay between Coriolis, buoyancy force. \color{black}

This phenomenology is supported by the scaling for the heat flux 
$Nu^{*} = (0.0046\pm0.0005)\times (Ra_{q}^*)^{0.26\pm0.04}$, which is also found in 
this regime.\\ \color{black}

Fifth, the geometry of the flow in the vicinity of the TC lateral boundary 
confirms that the Taylor-Proudman constraint is practically not broken there. 
On the other hand, for intense enough convection, the likely Stewartson layers 
that develop along this boundary diffuse a small part of the momentum 
generated inside the TC towards regions outside it.\\

Since our work has been largely motivated by the study of planetary 
cores, it is tempting to try and gain insight into their dynamics from these 
conclusions. This endeavour, however, meets several important obstacles. 
The first is the difference in the properties of the
working fluids. Recently revised estimates
of outer core thermal conductivity \citep{de2012electrical, pozzo2012thermal}
suggest that the Prandtl number
could have a low value $Pr \sim 10^{-2}$. The onset of rotating convection 
at such $Pr$ would be oscillatory even in the plane configuration. 
Nevertheless, it is still 
reasonable to expect that confinement within the TC induced by the virtual 
boundaries raised by the TP constraint reshapes convection in a similar way 
as it does at the moderate Prandtl numbers considered in this paper. 
This view is supported by the linear stability 
analysis of convection in a rotating cylinder at low Prandtl numbers by 
\cite{goldstein1994_jfm}, which shows that the critical modes are subject to 
both precession and oscillations.
Furthermore, convection in the Earth core is also compositional in nature, 
rather than only thermal. The corresponding Schmidt numbers are in a range 
comparable to the thermal Prandtl numbers of water and acid, 
for which the onset of convection would not be oscillatory.
Still, a deeper understanding of 
convection in a TC at low Prandtl number would require liquid metal experiments 
in a configuration similar to the present paper. \color{black}This would clarify 
the question of whether the oscillatory modes at the onset of low-$Pr$ convection are 
as robust to a change of boundary condition at the lateral boundary of the cylinder 
as we found the modes of moderate-$Pr$ convection to be.\\
Secondly, several examples of nonmagnetic spherical shell convection driven by 
moderately supercritical
convection of thermal \citep{sreenivasan2006azimuthal} or double-diffusive
(e.g. \citep{trumper2012numerical}) origin within the TC show an 
ensemble of thin viscously
controlled plumes with no indication of any coherent $z$-vorticity.
By contrast, coherent anticyclonic vorticity within the TC
has been noted in rotating dynamo
simulations \citep{glatzmaier1995_nat,sreenivasan2006azimuthal,schaeffer2017_gji}, 
which points to a possible role of the TC magnetic field in generating
strong polar vortices in moderately supercritical convection. \color{black}
That said, the central retrograde vortex noted in our experimental TC occurs
at $R_c \sim 10$, which might correspond to values of
$R_c \sim 50$--$100$ outside the 
spherical shell TC, given that convection inside
the TC sets in at a Rayleigh number much higher than outside
it \citep{jones2007thermal}. This strongly driven regime has not been
adequately explored, at least in the
computationally demanding regime of low $E$. Since
the criticality of convection within Earth's TC is not
well constrained, a comparison of
the convection pattern at $R_c \sim 10$
within our experimental TC with that in
the spherical shell TC would help ascertain whether the
two patterns bear any resemblance to each other.\\
Finally, understanding the role of the Lorentz forces within 
the Earth's TC 
is not as straightforward as one might imagine. 
With a uniform magnetic field, it is known that the Lorentz forces 
favour large-scale magnetically controlled plumes in a plane layer
\citep[e.g.][]{Aujogue_pof14}. However,
geodynamo simulations suggest that the mean field within the TC has
severe lateral and axial inhomogeneities, whose effect on the
width of convection plumes needs to be understood. \textcolor{black}{Further, 
the nature of the free shear layers at the cylinder boundary may change 
as a consequence of the constraints that the Lorentz force has to 
satisfy to ensure sufficient smoothness of the velocity 
field \citep{livermore2012_jmp,hollerbach1994_pf}}.
It is hoped that magnetoconvection experiments 
in a TC configuration at moderate Prandtl number 
(which our apparatus is 
designed to perform) could provide us with a 
somewhat more realistic picture of 
convection within the TC than rotating convection alone.\\

\color{black}K.A., A.P., and B.S. acknowledge support from the
Leverhulme Trust for this project (Research Grant No. RPG-
2012-456), and A.P. acknowledges support from the Royal
Society under the Wolfson Research Merit Award scheme.
K.A. acknowledges the support of the Royal Astronomical
Society. The authors are indebted to the LNCMI and its
technical and academic staff for the quality and effectiveness
of their support. Finally, the authors are indebted to the anonymous 
referees for their constructive remarks.
\color{black}

\bibliography{scholar2}

\begin{thebibliography}{48}
\expandafter\ifx\csname natexlab\endcsname\relax\def\natexlab#1{#1}\fi

\bibitem[Aubert(2005)]{aubert2005steady}
{\sc Aubert, J.} 2005 Steady zonal flows in spherical shell dynamos. {\em
  Journal of Fluid Mechanics\/} {\bf 542}, 53--67.

\bibitem[Aubert {\em et~al.\/}(2001)Aubert, Brito, Nataf, Cardin \&
  Masson]{aubert2001systematic}
{\sc Aubert, J., Brito, D., Nataf, H.-C., Cardin, P. \& Masson, J.-P.} 2001 A
  systematic experimental study of rapidly rotating spherical convection in
  water and liquid gallium. {\em Physics of the Earth and Planetary
  Interiors\/} {\bf 128}~(1), 51--74.

\bibitem[Aujogue {\em et~al.\/}(2016)Aujogue, Poth\'erat, Bates, Debray \&
  Sreenivasan]{Aujogue_rsi16}
{\sc Aujogue, K., Poth\'erat, A., Bates, I., Debray, F. \& Sreenivasan, B.}
  2016 Little earth experiment: An instrument to model planetary cores. {\em
  Review of Scientific Instruments\/} {\bf 87}~(8), 084502.

\bibitem[Aujogue {\em et~al.\/}(2015)Aujogue, Poth\'erat \&
  Sreenivasan]{Aujogue_pof14}
{\sc Aujogue, K., Poth\'erat, A. \& Sreenivasan, B.} 2015 Onset of plane layer
  magnetoconvection at low ekman number. {\em Physics of Fluids\/} {\bf
  27}~(10), 106602.

\bibitem[Aurnou(2007)]{aurnou2007planetary}
{\sc Aurnou, J.} 2007 Planetary core dynamics and convective heat transfer
  scaling. {\em Geophysical and Astrophysical Fluid Dynamics\/} {\bf
  101}~(5-6), 327--345.

\bibitem[Aurnou {\em et~al.\/}(2003)Aurnou, Andreadis, Zhu \&
  Olson]{aurnou2003experiments}
{\sc Aurnou, J., Andreadis, S., Zhu, L. \& Olson, P.} 2003 Experiments on
  convection in {E}arth's core tangent cylinder. {\em Earth Planet. Sci.
  Lett.\/} {\bf 212}~(1), 119--134.

\bibitem[Aurnou \& Olson(2001)]{aurnou2001experiments}
{\sc Aurnou, J.~M. \& Olson, P.} 2001 Experiments on {R}ayleigh--{B}{\'e}nard
  convection, magnetoconvection and rotating magnetoconvection in liquid
  gallium. {\em J. Fluid Mech.\/} {\bf 430}, 283--307.

\bibitem[Busse(1970)]{busse1970thermal}
{\sc Busse, F.~H.} 1970 Thermal instabilities in rapidly rotating systems. {\em
  J. Fluid Mech.\/} {\bf 44}~(03), 441--460.

\bibitem[Cardin \& Olson(1994)]{cardin1994chaotic}
{\sc Cardin, P. \& Olson, P.} 1994 Chaotic thermal convection in a rapidly
  rotating spherical shell: consequences for flow in the outer core. {\em
  Physics of the earth and planetary interiors\/} {\bf 82}~(3-4), 235--259.

\bibitem[Chandrasekhar(1961)]{chandrasekhar1961hydrodynamic}
{\sc Chandrasekhar, S.} 1961 {\em Hydrodynamic and hydromagnetic stability\/}.
  Clarendon Press, Oxford.

\bibitem[Cheng \& Aurnou(2016)]{cheng2016_espl}
{\sc Cheng, J.S. \& Aurnou, J.M.} 2016 Tests of diffusion-free scaling
  behaviors in numerical dynamo datasets. {\em Earth and Planetary Science
  Letters\/} {\bf 436}, 121 -- 129.

\bibitem[Cheng {\em et~al.\/}(2015)Cheng, Stellmach, Ribeiro, Grannan, King \&
  Aurnou]{cheng2015_gji}
{\sc Cheng, JS, Stellmach, S, Ribeiro, A, Grannan, A, King, EM \& Aurnou, JM}
  2015 Laboratory-numerical models of rapidly rotating convection in planetary
  cores. {\em Geophysical Journal International\/} {\bf 201}~(1), 1--17.

\bibitem[Christensen(2002)]{christensen2002zonal}
{\sc Christensen, U.R.} 2002 Zonal flow driven by strongly supercritical
  convection in rotating spherical shells. {\em Journal of Fluid Mechanics\/}
  {\bf 470}, 115--133.

\bibitem[Christensen \& Aubert(2006)]{christensen2006scaling}
{\sc Christensen, U.R. \& Aubert, J.} 2006 Scaling properties of
  convection-driven dynamos in rotating spherical shells and application to
  planetary magnetic fields. {\em Geophysical Journal International\/} {\bf
  166}~(1), 97--114.

\bibitem[Clune \& Knoblauch(1993)]{clune1993_pre}
{\sc Clune, T. \& Knoblauch, E.} 1993 Pattern selection in rotating convection
  with experimental boundary conditions. {\em Phys. Rev. E\/} {\bf 47}~(4),
  2536--2540.

\bibitem[Cui \& Street(2001)]{cui2001_jfm}
{\sc Cui, Anquing \& Street, Robert~L.} 2001 Large-eddy simulation of turbulent
  rotating convective flow development. {\em Journal of Fluid Mechanics\/} {\bf
  447}, 53¿84.

\bibitem[Curbelo {\em et~al.\/}(2014)Curbelo, Lopez, Mancho \&
  Marques]{curbelo2014_pre}
{\sc Curbelo, Jezabel, Lopez, Juan~M., Mancho, Ana~M. \& Marques, Francisco}
  2014 Confined rotating convection with large prandtl number: Centrifugal
  effects on wall modes. {\em Phys. Rev. E\/} {\bf 89}, 013019.

\bibitem[Ecke {\em et~al.\/}(1992)Ecke, Zhong \& Knobloch]{ecke1992hopf}
{\sc Ecke, R.E., Zhong, F. \& Knobloch, E.} 1992 Hopf bifurcation with broken
  reflection symmetry in rotating rayleigh-b{\'e}nard convection. {\em EPL
  (Europhysics Letters)\/} {\bf 19}~(3), 177.

\bibitem[Gastine {\em et~al.\/}(2016)Gastine, Wicht \& Aubert]{gastine2016_jfm}
{\sc Gastine, Thomas, Wicht, Johannes \& Aubert, Julien} 2016 Scaling regimes
  in spherical shell rotating convection. {\em Journal of Fluid Mechanics\/}
  {\bf 808}, 690--732.

\bibitem[Gastine {\em et~al.\/}(2015)Gastine, Wicht \& Aurnou]{gastine2015_jfm}
{\sc Gastine, Thomas, Wicht, Johannes \& Aurnou, Jonathan~M.} 2015 Turbulent
  rayleigh¿bénard convection in spherical shells. {\em Journal of Fluid
  Mechanics\/} {\bf 778}, 721--764.

\bibitem[Glatzmaiers \& Roberts(1995)]{glatzmaier1995_nat}
{\sc Glatzmaiers, G.A. \& Roberts, P.H.} 1995 A three-dimensional
  self-consistent computer simulation of a geomagnetic field reversal. {\em
  Nature\/} {\bf 377}, 203 -- 209.

\bibitem[Goldstein {\em et~al.\/}(1994)Goldstein, Knoblauch, Mercader \&
  Net]{goldstein1994_jfm}
{\sc Goldstein, H.F., Knoblauch, E., Mercader, I. \& Net, M.} 1994 Convection
  in a rotating cylinder. part 2. linear theory for low prandtl numbers. {\em
  Journal of Fluid Mechanics\/} {\bf 262}, 293--324.

\bibitem[Goldstein {\em et~al.\/}(1993)Goldstein, Knobloch, Mercader \&
  Net]{goldstein1993convection}
{\sc Goldstein, H.F., Knobloch, E., Mercader, I. \& Net, M.} 1993 Convection in
  a rotating cylinder. part 1 linear theory for moderate prandtl numbers. {\em
  Journal of Fluid Mechanics\/} {\bf 248}, 583--604.

\bibitem[Grossmann \& Lohse(2000)]{grossmann2000_jfm}
{\sc Grossmann, S. \& Lohse, D.} 2000 Scaling in thermal convection: a unifying
  theory. {\em Journal of Fluid Mechanics\/} {\bf 407}, 27¿56.

\bibitem[Hollerbach(1994)]{hollerbach1994_pf}
{\sc Hollerbach, Rainer} 1994 Imposing a magnetic field across a
  nonaxisymmetric shear layer in a rotating spherical shell. {\em Physics of
  Fluids\/} {\bf 6}~(7), 2540--2544.

\bibitem[Horn \& Shishkina(2014)]{horn2014_pf}
{\sc Horn, S. \& Shishkina, O.} 2014 Rotating non-oberbeck¿boussinesq
  rayleigh¿b\'enard convection in water. {\em Phys. Fluids.\/} {\bf 26},
  055111.

\bibitem[Hulot {\em et~al.\/}(2002)Hulot, Eymin, Langlais, Mandea \&
  Olsen]{hulot2002small}
{\sc Hulot, Gauthier, Eymin, C{\'e}line, Langlais, Beno{\^\i}t, Mandea, Mioara
  \& Olsen, Nils} 2002 Small-scale structure of the geodynamo inferred from
  oersted and magsat satellite data. {\em Nature\/} {\bf 416}~(6881), 620--623.

\bibitem[Jacobs \& Ivey(1998)]{jacobs1998_jfm}
{\sc Jacobs, P. \& Ivey, G.~N.} 1998 The influence of rotation on shelf
  convection. {\em Journal of Fluid Mechanics\/} {\bf 369}, 23¿48.

\bibitem[Jones(2007)]{jones2007thermal}
{\sc Jones, C.A.} 2007 Thermal and compositional convection in the outer core.
  {\em Treatise in Geophysics, Core Dynamics\/} {\bf 8}, 131--185.

\bibitem[Julien {\em et~al.\/}(2012)Julien, Knobloch, Rubio \&
  Vasil]{julien2012_prl}
{\sc Julien, Keith, Knobloch, Edgar, Rubio, Antonio~M \& Vasil, Geoffrey~M}
  2012 Heat transport in low-rossby-number rayleigh-b{\'e}nard convection. {\em
  Physical review letters\/} {\bf 109}~(25), 254503.

\bibitem[King {\em et~al.\/}(2012)King, Stellmach \& Aurnou]{king2012_jfm}
{\sc King, E.~M., Stellmach, S. \& Aurnou, J.~M.} 2012 Heat transfer by rapidly
  rotating rayleigh¿bénard convection. {\em Journal of Fluid Mechanics\/} {\bf
  691}, 568¿582.

\bibitem[de~Koker {\em et~al.\/}(2012)de~Koker, Steinle-Neumann \&
  Vl{\v{c}}ek]{de2012electrical}
{\sc de~Koker, N., Steinle-Neumann, G. \& Vl{\v{c}}ek, V.} 2012 Electrical
  resistivity and thermal conductivity of liquid fe alloys at high p and t, and
  heat flux in earthÕs core. {\em Proceedings of the National Academy of
  Sciences\/} {\bf 109}~(11), 4070--4073.

\bibitem[Kraichnan(1962)]{kraichnan1962_pf}
{\sc Kraichnan, R.~H.} 1962 Turbulent thermal convection at arbitrary prandtl
  number. {\em Phys. Fluids 5\/} p. 1374¿1389.

\bibitem[Kunnen {\em et~al.\/}(2010)Kunnen, Geurts \& Clerx]{kunnen2010_jfm}
{\sc Kunnen, R. P.~J., Geurts, B.~J. \& Clerx, H. J.~H.} 2010 Experimental and
  numerical investigation of turbulent convection in a rotating cylinder. {\em
  Journal of Fluid Mechanics\/} {\bf 642}, 445¿476.

\bibitem[Kunnen {\em et~al.\/}(2011)Kunnen, Stevens, Overkamp, Sun, van Heijst
  \& Clercx]{kunnen2011_jfm}
{\sc Kunnen, Rudie P.~J., Stevens, Richard J. A.~M., Overkamp, Jim, Sun, Chao,
  van Heijst, GertJan~F. \& Clercx, Herman J.~H.} 2011 The role of stewartson
  and ekman layers in turbulent rotating rayleigh¿bénard convection. {\em
  Journal of Fluid Mechanics\/} {\bf 688}, 422¿442.

\bibitem[Livermore {\em et~al.\/}(2017)Livermore, Hollerbach \&
  Finlay]{livermore2017_natgeo}
{\sc Livermore, P.W., Hollerbach, R. \& Finlay, C.} 2017 An accelerating
  high-latitude jet in earth's core. {\em Nat. Geo.\/} {\bf 10}, 62--69.

\bibitem[Livermore \& Hollerbach(2012)]{livermore2012_jmp}
{\sc Livermore, Philip~W. \& Hollerbach, Rainer} 2012 Successive elimination of
  shear layers by a hierarchy of constraints in inviscid spherical-shell flows.
  {\em Journal of Mathematical Physics\/} {\bf 53}~(7), 073104.

\bibitem[Marques {\em et~al.\/}(2007)Marques, Mercader, Batiste \&
  Lopez]{marques2007_jfm}
{\sc Marques, F., Mercader, I, Batiste, O. \& Lopez, J.~M.} 2007 Centrifugal
  effects in rotating convection: axisymmetric states and three-dimensional
  instabilities. {\em Journal of Fluid Mechanics\/} {\bf 580}, 303¿318.

\bibitem[Maxworthy \& Narimousa(1994)]{maxworthy1994_jpo}
{\sc Maxworthy, T. \& Narimousa, S.} 1994 Unsteady turbulent convection into a
  homogeneous rotating fluid with oceanic applications. {\em J. Phys. Ocean.\/}
  {\bf 24}, 865--887.

\bibitem[Pozzo {\em et~al.\/}(2012)Pozzo, Davies, Gubbins \&
  Alfe]{pozzo2012thermal}
{\sc Pozzo, M., Davies, C., Gubbins, D. \& Alfe, D.} 2012 Thermal and
  electrical conductivity of iron at earth's core conditions. {\em Nature\/}
  {\bf 485}~(7398), 355--358.

\bibitem[Schaeffer {\em et~al.\/}(2017)Schaeffer, Jault, Nataf \&
  Fournier]{schaeffer2017_gji}
{\sc Schaeffer, N., Jault, D., Nataf, H.-C. \& Fournier, A.} 2017 Turbulent
  geodynamo simulations: a leap towards earth¿s core. {\em Geophys. J. Int.\/}
  p. in press.

\bibitem[Schubert \& Soderlund(2011)]{schubert2011_pepi}
{\sc Schubert, G. \& Soderlund, K.M.} 2011 Planetary magnetic fields:
  Observations and models. {\em Physics of the Earth and Planetary Interiors\/}
  {\bf 187}~(3), 92--108.

\bibitem[Sreenivasan \& Jones(2006)]{sreenivasan2006azimuthal}
{\sc Sreenivasan, B. \& Jones, C.~A.} 2006 Azimuthal winds, convection and
  dynamo action in the polar regions of planetary cores. {\em Geophys.
  Astrophys. Fluid Dyn.\/} {\bf 100}~(4-5), 319--339.

\bibitem[Stewartson(1957)]{stewartson1957}
{\sc Stewartson, K} 1957 On almost rigid rotations. {\em Journal of Fluid
  Mechanics\/} {\bf 3}~(1), 17--26.

\bibitem[Sumita \& Olson(2003)]{sumita2003experiments}
{\sc Sumita, I. \& Olson, P.} 2003 Experiments on highly supercritical thermal
  convection in a rapidly rotating hemispherical shell. {\em Journal of Fluid
  Mechanics\/} {\bf 492}, 271--287.

\bibitem[Tr{\"u}mper {\em et~al.\/}(2012)Tr{\"u}mper, Breuer \&
  Hansen]{trumper2012numerical}
{\sc Tr{\"u}mper, T., Breuer, M. \& Hansen, U.} 2012 Numerical study on
  double-diffusive convection in the earthÕs core. {\em Physics of the Earth
  and Planetary Interiors\/} {\bf 194}, 55--63.

\bibitem[Zhang \& Liao(2009)]{zhang2009_jfm}
{\sc Zhang, K. \& Liao, X.} 2009 The onset of convection in rotating circular
  cylinders with experimental boundary conditions. {\em J. Fluid Mech.\/} {\bf
  622}, 63¿73.

\bibitem[Zhong {\em et~al.\/}(1993)Zhong, Ecke \& Steinberg]{zhong1993rotating}
{\sc Zhong, F., Ecke, R.E. \& Steinberg, V.} 1993 Rotating rayleigh--b{\'e}nard
  convection: asymmetric modes and vortex states. {\em Journal of Fluid
  Mechanics\/} {\bf 249}, 135--159.

\end{thebibliography}
\bibliographystyle{jfm}

\end{document}